\documentclass[manuscript]{aastex631}
\submitjournal{arXiv}

\shorttitle{Geochemical Hazard Assessment of Martian Regolith}
\shortauthors{Anjum \& Bart}
\graphicspath{{./}{figures/}}

\usepackage{amsmath} 
\usepackage{acronym} 
\usepackage{xpatch}
\xpatchcmd{\thebibliography}{\twocolumngrid}{}{}{}

\begin{document}

\title{{Geochemical Hazard Assessment of Martian Regolith for Future Human Exploration}}

\correspondingauthor{Mavia Anjum}
\email{mav.mavia14@gmail.com}

\author[0009-0005-2773-4172]{Mavia Anjum}
\affiliation{University of Idaho \\
Department of Physics\\
875 Perimeter Dr., MS 0903 \\
Moscow, Idaho, 83844-0903, USA}

\author[0000-0002-4768-2972]{Gwendolyn D. Bart}
\affiliation{University of Idaho \\
Department of Physics\\
875 Perimeter Dr., MS 0903 \\
Moscow, Idaho, 83844-0903, USA}

\begin{abstract}

As human missions to Mars move from concept to planning to reality, a systematic quantitative health risk assessment of martian regolith exposure has become critically important. This study presents a comprehensive multi-element, multi-pathway health hazard analysis of martian regolith for a 70~kg adult astronaut on an 18-month surface mission, using bulk silicate Mars geochemical data and Earth Upper Continental Crust (UCC) reference values as baseline comparators. We computed \acf{ADD}, \acf{HQ}, \acf{HI}, \acf{EnFac}, \acf{ERF}, and \acf{ILCR} across all three exposure pathways (oral ingestion, inhalation, and dermal contact) for toxic and heavy metals, under four filtration scenarios (0\%, 25\%, 50\%, 95\%). Results demonstrate that Cr and Co represent the most critical non-carcinogenic hazards, exceeding the regulatory threshold (\ac{HI} $>$ 1) even at 50\% filtration, while Cr also poses the most significant carcinogenic risk, near the $10^{-4}$ regulatory threshold level. At least 92\% regolith filtration efficiency is required to reduce Cr to acceptable non-cancer hazard levels. Nickel (Ni) exceeds the acceptable cancer risk threshold of $10^{-6}$ at all filtration levels below 95\%. No element other than Cr and Co exceeds \ac{HI}~=~1 in the unfiltered scenario, although Fe, Ni and Mn approach concerning levels. These results establish that a minimum 95\% \ac{HEPA}-grade filtration efficiency combined with active chemical sorption is required for acceptable Cr and Co short term exposure and Cr, Ni long term exposure management. This study provides the first element resolved, pathway-specific quantitative risk framework directly applicable to habitat air quality standards, EVA suit filtration specifications, and pre-mission biomonitoring protocols for crewed Mars surface operations.

\end{abstract}



\section{Introduction} \label{sec:intro}

The potential of human exploration of Mars has evolved from a long-term dream to a near-term possibility, with the \ac{NASA}'s Artemis program, commercial entities such as SpaceX, and international partners actively developing mission architectures for crewed Mars surface operations. Despite remarkable advances in propulsion, life support, and autonomous systems, the chemical and toxicological threats posed by martian regolith are still one of the unexplored risks to astronaut health. In contrast to radiation exposure, which has been extensively simulated using data from the \ac{MSL}'s \ac{RAD} and other instruments, the geochemical danger from regolith exposure has received less systematic quantitative assessment \citep{patel2020, cucinotta2013}.

Martian regolith is a chemically complex, globally distributed, fine-grained material derived primarily from the basaltic crust of Mars through aeolian, impact, and aqueous weathering processes. Although the size distribution of martian dust is not uniform but, it is considered to be ~ 2-3~$\mu$m for suspended dust fractions and that makes it highly respirable and capable of deep lung deposition \citep{cucinotta2013, wang2021}. Martian dust is less abrasive than lunar dust, due to higher erosion rates, but still a credible threat to astronauts. It is electrostatic, magnetic and chemically volatile, composed of irregularly shaped grains with smoothed vertices. The average particle diameter can range from 2-8~$\mu$m during dust events \citep{wang2021}. Several toxic and potentially carcinogenic elements have been identified in martian regolith or inferred from bulk geochemistry: perchlorates \citep{davila2013}; crystalline silica capable of inducing silicosis \citep{hoy2020,kayama2018}; nanophase iron oxides that can generate reactive oxygen species \citep{bell1995}; and trace levels of chromium (Cr), cobalt (Co), nickel (Ni), beryllium (Be), arsenic (As), and cadmium (Cd), which are classified by the \ac{EPA} as known carcinogens \citep{taylor1985, hadjiliadis2012, banfalvi2011}. Their concentration and potential health risks are vital to study for planning future Mars missions and astronaut safety.

Standard \ac{EPA} risk assessment methodology \citep{phillips2013epa} distinguishes three primary routes of soil/regolith exposure: (1) incidental oral ingestion of particles; (2) inhalation of resuspended particulate matter; and (3) dermal absorption through skin contact with contaminated surfaces \citep{anjum2024i}. In the context of Mars surface and future manned missions, all three pathways are relevant, but with important modifications relative to terrestrial settings. The near-vacuum martian atmosphere (mean pressure ~600~Pa, approximately 0.6\% of Earth's sea-level pressure) means dust behaviour is dominated by electrostatic forces rather than aerodynamic settling; martian gravity ($g_{mars}=0.376 g_{earth}$) reduces settling velocity, maintaining higher sustained dust concentrations than equivalent terrestrial conditions \citep{calle2011, rahman2023}. This implies that a comprehensive health and toxicology framework should be established to evaluate the risks to astronauts that are unique to the martian environment.
The \ac{EPA} \ac{HHRA} toxicological framework has been widely applied in diverse terrestrial contamination scenarios (e.g. for soil, water, and food) \citep{anjum2024m, anjum2024i, mao2019}. But this toxicological model has not yet been applied for martian regolith. 

\citet{wang2021} analyzed the clinical consequences of martian dust for astronauts and concluded that expected effects range from asymptomatic to life-threatening, with the respiratory system carrying the greatest burden of chronic disease risk. However, that analysis did not statistically integrate the toxicological model which helps to calculate risk metrics. Likewise, the \ac{NASA} Human Research Program \citep{patel2020} has designated dust-induced lung disease as a gap knowledge area and recognizes dust as a closed-environment danger requiring active mitigation \citep{nasahuman}. Therefore, the toxicological consequences of martian regolith on human health must be investigated for future manned missions to Mars. This study aims to do that by performing a comprehensive multi-element, multi-pathway quantitative health risk assessment of martian regolith reported to date. We use the Mars elemental dataset compiled and reassessed by \citet{taylor2013}, which gathered the data from 60+ martian meteorites, the Mars Odyssey \ac{GRS} and several lander and rover \ac{APXS} datasets. We compare the results with the Earth's \ac{UCC} composition \citep{wedepohl1995}, improve the conventional \ac{EPA} risk model for Mars and calculate metrics for several toxic elements, effective reduction of the risk by modelling four filtration scenarios, and also give Monte Carlo uncertainty for risk threshold analysis. The outputs are expected to directly impact mission design requirements, particularly airlock and habitat air filtration parameters, \ac{EVA} suit design, and medical countermeasure strategies.

\newpage
\section{Methodology}\label{sec:method}
The methodology follows the \ac{EPA} \ac{RAGS} framework \citep{eparags}. We have made adaptations in this model for the Mars surface environment. The workflow comprises four stages: (1) elemental data compilation; (2) Mars-specific exposure parameter derivation; (3) dose and risk metric computation across filtration scenarios. Each stage is described in detail below, with explicit justification for all parameter choices that deviate from terrestrial defaults set by the \ac{EPA}.

\subsection{Elemental Data Sources}
Mars regolith elemental concentrations were taken from the revised bulk silicate Mars model of \citet{taylor2013}, which synthesized elemental data from $60+$ martian meteorites, the Mars Odyssey \ac{GRS} dataset and multiple lander and rover \ac{APXS} measurements. This dataset provides concentrations for 57 elements with $2\sigma$ uncertainties, making it the most comprehensive peer-reviewed Mars bulk composition currently available. For vanadium (V), not tabulated in \citet{taylor2013}, we estimated a bulk silicate Mars concentration of 80~ppm by scaling CI~chondrite \citep{wasson1974} vanadium abundance (57~ppm) using the refractory enrichment factor ($1.4\,\times$\,CI) implied by the martian elemental model of \citet{taylor2013}. Earth UCC reference values were taken from \citet{wedepohl1995}. This dataset was chosen over \citet{taylor1985} because \citet{wedepohl1995} is based on actual rock analyses rather than a seismically-derived model and is widely used as the reference for enrichment calculations. Measurement uncertainty for each element was characterized prior to Monte Carlo analysis. The published $2\sigma$ uncertainties \citep{taylor2013} were used directly.

\subsection{Enrichment Factor and Ecological Risk Factor Calculation}

The enrichment of different elements in martian regolith with respect to the \acf{UCC} of Earth is calculated using the following relation:
\begin{equation}
 Enrichment Factor (EnFac)~=~\dfrac{\frac{C_{Element, Mars}}{C_{Al, Mars}}}{\frac{C_{Element, Earth\,UUC}}{C_{Al, Earth UUC}}}
\end{equation}
Here, $C_{Element, Mars}$ is the concentration of an element in martian regolith, $C_{Al, Mars}$ is the concentration of Al on the surface of Mars. Whereas $C_{Element, Earth\,UUC}$ is the elemental composition of that particular element on the UCC of the Earth and $C_{Al, Earth UUC}$ is the \ac{UCC} concentration of Al. Generally, in environmental studies enrichment factor is calculated as site specific value of an element with respect to its value in \ac{UCC} normalized to Aluminium concentration. Here, we modify this approach and present the formula for calculating Enrichment Factor for martian regolith with respect to Earth’s crust.
Ecological risk factor first given by \citet{hakanson1980} is used to calculate the pollution level and risk posed by the toxic and heavy elements to the ecology of a region. For the martian regolith ecological risk factor is calculated using the following relation:
\begin{equation}
    Ecological Risk Factor (ERF) =   \dfrac{C_{Mars}}{C_{EarthUCC}} \times TRF
\end{equation}
In this formula, $C_{Mars}$ is the concentration of an element in the martian regolith, $C_{EarthUCC}$ is the upper continental crust value of that element on Earth and \ac{TRF} represents the toxicity of a particular element. \ac{TRF}s for canonical metals (e.g. Cd~=~30, Hg~=~40, As~=~10) were taken from the original Håkanson framework. For elements not in the original model (e.g. Co), \ac{TRF}s were allocated based on generally reported values in follow-up ecological risk studies with Co assigned a value of 5 reflecting moderate toxicity \citep{anjum2024m}. For the elements, which does not have a toxic response factor in the literature, were assigned a \ac{TRF} value of 1. The base parameters used to calculate \ac{EnFac} and \ac{ERF} employed in this study will provide a baseline for the risk assessment of martian regolith for future works. 

\subsection{Exposure Parameter Derivation and Mars-Specific Adjustments}

For the health hazards assessment, a standard 70~kg adult astronaut was modelled. Mission duration was set to 18 months (\ac{ED}~=~1.5~years; 547~days), representing the nominal surface stay in current Mars Design Reference Architecture scenarios. \ac{ExFreq} was set to 365~days/year, assuming continuous habitation with no exposure-free periods. \ac{RIng}) was set to 100~mg/day. This exceeds the standard \ac{EPA} adult value of 50~mg/day by a factor of 2, which is justified on three grounds specific to the Mars \ac{EVA} environment. First, regolith tracking into airlock environments creates persistent surface contamination that is ingested during suit doffing and subsequent activities. Second, the charged, adhesive nature of martian nanophase iron oxide particles will increases adherence to glove surfaces and subsequent hand-to-mouth transfer \citep{banin1993}.  \ac{IR} was set to 20 m$^3$/day, which is the standard \ac{EPA} adult value \citep{choi2025}. The atmospheric dust concentration inside \ac{EVA} suits and habitats was modelled using the following parameters, K~=~$1.0 \times 10^{-6} m^{-1}$ is the standard terrestrial soil-to-air resuspension factor, and \ac{MDF}~=~2.5. The \ac{MDF} value accounts for: (a) reduced gravitational settling ($g_{mars}=0.376 g_{earth}$) which elevates suspended particle concentrations by approximately a factor of 2–3 for equivalent surface disturbance; (b) the absence of vegetation and surface moisture that damp dust resuspension on Earth; and (c) the elevated frequency and intensity of global and regional dust storms on Mars. The suit filtration efficiency factor of 0.95 reflects real-world \ac{EVA} filter performance accounting for edge-leak and filter loading over time. Dermal exposure parameters were: exposed skin surface area \ac{SA}~=~3,300 cm$^2$ (\ac{EVA} glove gaps and face seal areas); \acf{AF}~=~0.07 mg/cm$^2$; and  \acf{ABSd}~=~0.001, representing strongly attenuated dermal penetration under full \ac{EVA} suit conditions. This \ac{ABSd} is 10$\times$ lower than the \ac{EPA} default of 0.01 for bare skin contact, justified by the physical barrier provided by \ac{EVA} suit materials. \acf{ATnc}~=~\ac{ED} $\times$ 365~=~547 days; \acf{ATc}~=~70 $\times$ 365~=~25,550 days. Geochemical uncertainty for health hazards was propagated using Monte Carlo simulation with N~=~10,000 draws per element. For each element, concentrations were sampled from log-normal distributions. Using the above parameters in the following formulae, various health hazard parameters were calculated to quantify non-carcinogenic and carcinogenic health hazards. 

\acf{ADD} via each pathway was calculated using the standard \ac{EPA} \ac{RAGS}  \citep{eparags} and \ac{EPA} \ac{IRIS} \citep{epairis} equations, after \citet{anjum2024i}:
\begin{equation}
ADD_{ing}~=~\dfrac{C \times RIng \times EF \times ED \times (1-Filt) \times 10^{-6}}{W \times AT}
\end{equation}
\begin{equation}
ADD_{inh} = \dfrac{Ceff \times K \times MDF \times IR \times (1-0.95 \times Filt)}{BW}  \end{equation}
\begin{equation}
ADD_{dermal}= \dfrac{C \times SA \times AF \times ABSd \times (1-Filt) \times 10^{-6}}{BW}
\end{equation}
where $C$ is elemental concentration in mg/kg (ppm), $Ceff = (1-f) \times C$, $Filt$ is the filtration efficiency (dimensionless, 0–1). The units of \ac{ADD} are mg/kg-day. The \ac{HQ} for each pathway was computed as $ HQ = ADD / RfD$, where \ac{RfD} is the \ac{EPA} \ac{IRIS} chronic oral Reference Dose (mg/kg-day). Reference Doses for all elements were sourced from the \ac{EPA} \ac{IRIS} database. Where \ac{IRIS} data were unavailable, we used Minimal Risk Levels from \ac{ATSDR} \citep{epairis,cdcatsdr}. The \ac{HI} was summed across the three pathways:
\begin{equation}
HI = HQ_{ing} + HQ_{inh} + HQ_{dermal}
\end{equation}

\ac{ILCR} was computed via the oral ingestion pathway:
\begin{equation}
ILCR = \dfrac{ADD_{ing} \times CSF \times ED \times 365}{ATc}
\end{equation}
where CSF is the \ac{EPA} \ac{IRIS} oral \ac{CSF} (per mg/kg-day). The restriction to the oral pathway for \ac{ILCR} reflects the limited availability of \ac{IUR} for all elements in the \ac{IRIS} database; this provides a conservative lower bound on carcinogenic risk. For Cr specifically, the \ac{IRIS} Cr(VI) oral \ac{CSF} of 0.5 (mg/kg-day)$^{-1}$ was applied, representing a worst-case scenario for the unknown Cr speciation on Mars.

\subsection{Filtration Scenario Design}
Four regolith filtration scenarios were evaluated: (1) Raw (0\% filtration): worst-case representing a complete filtration breach or unprotected access; (2) 25\% filtration: equivalent to basic dust mask performance; (3) 50\% filtration: moderate protection, such as a standard respirator; (4) 95\% filtration: representative of high-efficiency \acf{HEPA} filtration systems in pressurized habitats. Filtration was applied uniformly to all three exposure pathways in the base model, with the following physical justification: (a) oral ingestion reduction represents decontamination of airlock and suit surfaces before post-\ac{EVA} entry; (b) inhalation filtration is provided by the \ac{EVA} suit life support system; and (c) dermal filtration represents the reduction in surface contamination from suit washing protocols. In addition to the four discrete scenarios, continuous filtration efficacy was analyzed across the range 0–99\% to determine the minimum filtration efficiency required to achieve \ac{HI}~$<$~1 for each element whose \ac{HI} value exceeded 1. 
\newpage

\section{Results}\label{sec:results}

\subsection{Elemental Enrichment Factor and Ecological Risk Factor}

\subsubsection{Enrichment Factor}
\begin{figure}[htbp]
\begin{center}
\includegraphics[width=1.0\textwidth]{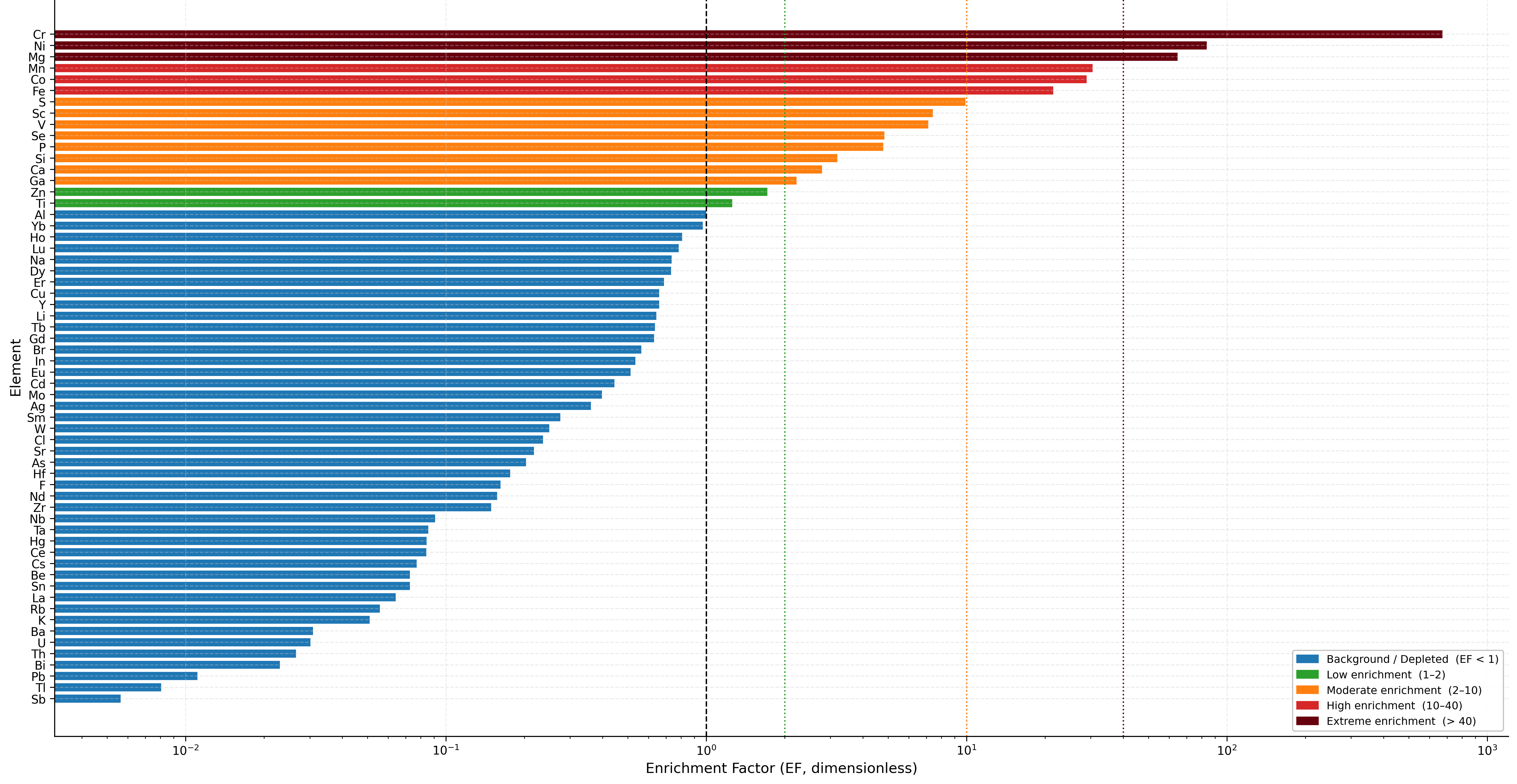}
\end{center}
\caption{Enrichment factor for martian elements normalized to Eaupper continental crust of Earth (Al).\label{fig1}}
\end{figure}

The \acf{EnFac} reveals a range spanning nearly five orders of magnitude across the elements analysed (Fig.~\ref{fig1}). The results could be divided into four geochemically distinct groups. The extreme enrichment group (\ac{EnFac} $>$ 40) contains only chromium (Cr; \ac{EnFac}~=~673), nickel (Ni; \ac{EnFac}~=~98) and Magnesium (Mg; \ac{EnFac}~=~65). Within this group, Chromium's extraordinarily high \ac{EnFac} arises from the combination of its near-CI chondritic abundance in the martian bulk composition and its strong depletion in the Earth \ac{UCC} caused by billions of years of differentiation into chromite-bearing lower crust and mantle \citep{taylor2013,wedepohl1995}. Specifically, (Cr)Mars~=~4,990 ppm versus (Cr)\ac{UCC}~=~35 ppm, yielding a very high concentration on Mars than Earth. The high enrichment of Ni in the martian regolith can be attributed to the lack of tectonic activity and accumulation of Nickel on the martian surface due to a high frequency of Fe-Ni rich impactors \citep{yen2006}. 

The high enrichment group (\ac{EnFac}~=~10-40) comprises, cobalt (Co; \ac{EnFac}=29), manganese (Mn; \ac{EnFac}~=~30), iron (Fe; \ac{EnFac}~=~22), and sulfur (S; \ac{EnFac}~=~10.5). Iron and manganese enrichment is well-established from the Mars Odyssey \ac{GRS} data \citep{boynton2007, lanza2014, treiman2023}, with martian FeO reaching ~18 weight\% compared to approximately 5.0 weight\% in the \ac{UCC} of Earth \citep{taylor2013}. Cobalt enrichment reflects its siderophile character and incomplete sequestration into the martian core \citep{boynton2007, lanza2014, treiman2023}. The moderate enrichment group (\ac{EnFac}~=~2–10) includes scandium (Sc; \ac{EnFac}~=~9.2), selenium (Se; \ac{EnFac}~=~4.8), phosphorus (P; \ac{EnFac}~=~4.5), and silicon (Si; \ac{EnFac}~=~4.1). The large depletion group (\ac{EnFac}~$<$~1) encompasses most of the elements, dominated by the \ac{LILE}: potassium (K; \ac{EnFac}~=~0.012), rubidium (Rb; \ac{EnFac}~=~0.022), cesium (Cs; \ac{EnFac}~=~0.016), barium (Ba; \ac{EnFac}~=~0.031), thorium (Th; \ac{EnFac}~=~0.026), uranium (U; \ac{EnFac}~=~0.029), and lead (Pb; \ac{EnFac}~=~0.011). This pattern is broadly consistent with the Wänke-Dreibus model \citep{wanke1988, wanke1994} for the bulk silicates on Mars. This model assumes that Cl-like refractory lithophile elemental ratios and predicts the depletion of elements which are moderately and highly volatile relative to the Cl abundances.

\subsubsection{Ecological Risk Factor}
\begin{figure}[htbp]
\begin{center}
\includegraphics[width=1.0\textwidth]{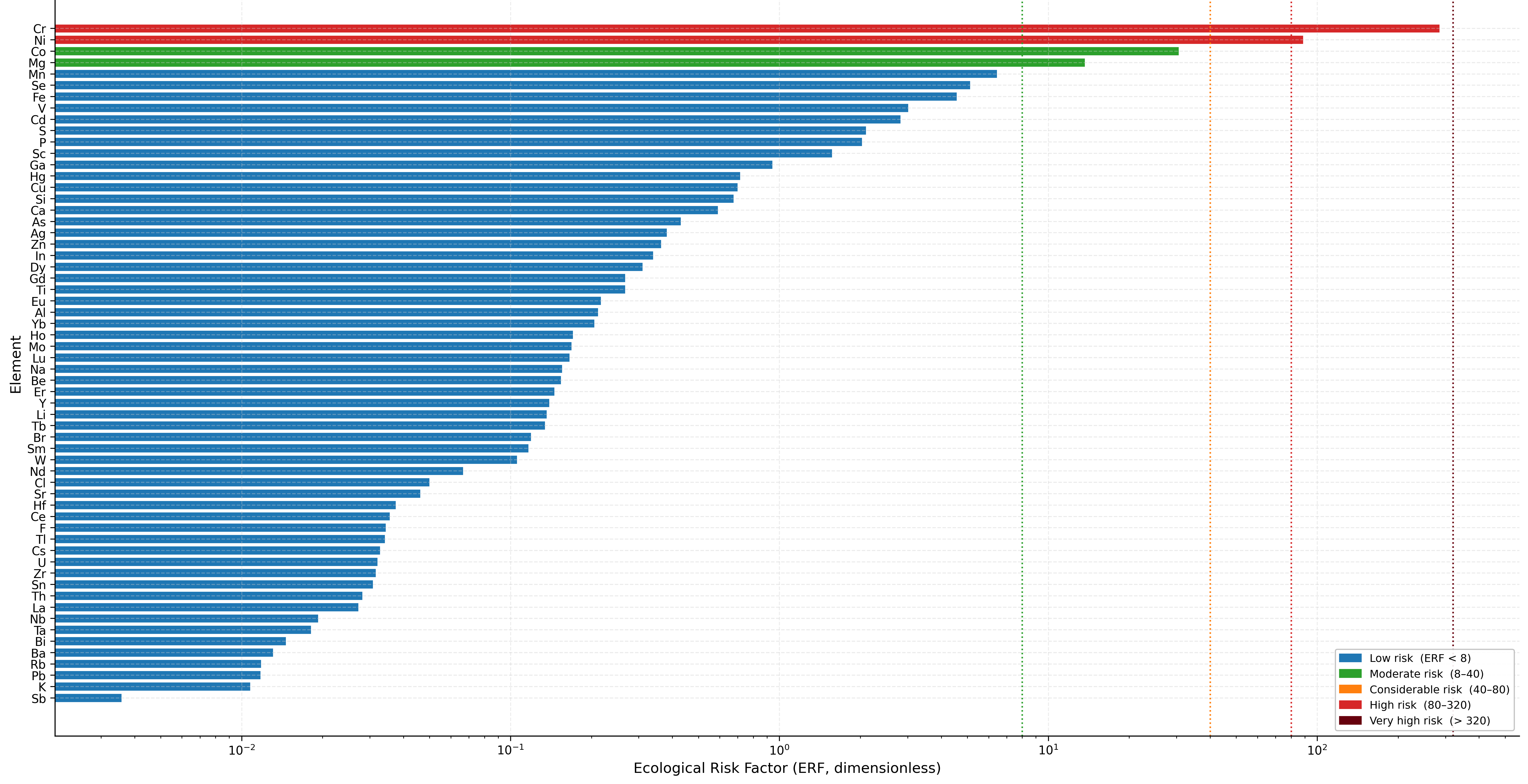}
\end{center}
\caption{Enrichment factor for martian elements normalized to upper continental crust of Earth (Al).\label{fig2}}
\end{figure}

The \acf{ERF} provides a toxicity-weighted geochemical risk index that integrates both abundance anomaly and biological potency (Fig.~\ref{fig2}). In the ecological risk analysis for martian regolith, Cr and Ni dominates with very high risk. These two elements stand in a class apart from the remainder of the dataset. In the moderate ecological risk range lie Co, Mg and Mn. The \ac{REE}, despite being strongly depleted relative to \ac{UCC}, cluster in the \ac{ERF} range 0.001–0.05 and pose no ecological or health significance in this context. Notably, As and Cd, both Group 1 \ac{IARC} carcinogens \citep{epairis}, rank well below Cr and Ni in the \ac{ERF} framework because their absolute concentrations in Mars are markedly lower than in Earth \ac{UCC} (As: 0.086 ppm Mars vs.\ 2.0 ppm \ac{UCC}; Cd: 9.6 ppb Mars vs.\ 102 ppb \ac{UCC}). 

The difference between the \ac{EnFac} and \ac{ERF} rankings is instructive: Mg and Fe have high \ac{EnFac} values, but their low \ac{TRF} (\ac{TRF}~=~1 for both) reduces their \ac{ERF}. In contrast, Ni's modest \ac{EnFac} is amplified by its \ac{TRF} of 5, which raises its \ac{ERF} over that of Co. This illustrates that prioritizing health risk based solely on geochemical abundance anomaly is insufficient; toxicological efficacy must also be considered. This finding is directly applicable to future human settlement of Mars. Without previous hazard control, the regolith cannot be viewed as an inert construction material or agricultural substrate, as increased Cr and Ni may represent the most important exposure pathway via inhalation, ingestion, and long-term dust contact. In a martian colonization scenario, \ac{ERF} may serve as a first-order screening technique for finding pollutants that could jeopardize crew health, life-support reliability, and the safe usage of local materials.

\subsection{Average Daily Dose}
\subsubsection{Mars vs.\ Earth UCC Comparison}
\begin{figure}[htbp]
\begin{center}
\includegraphics[width=1.0\textwidth]{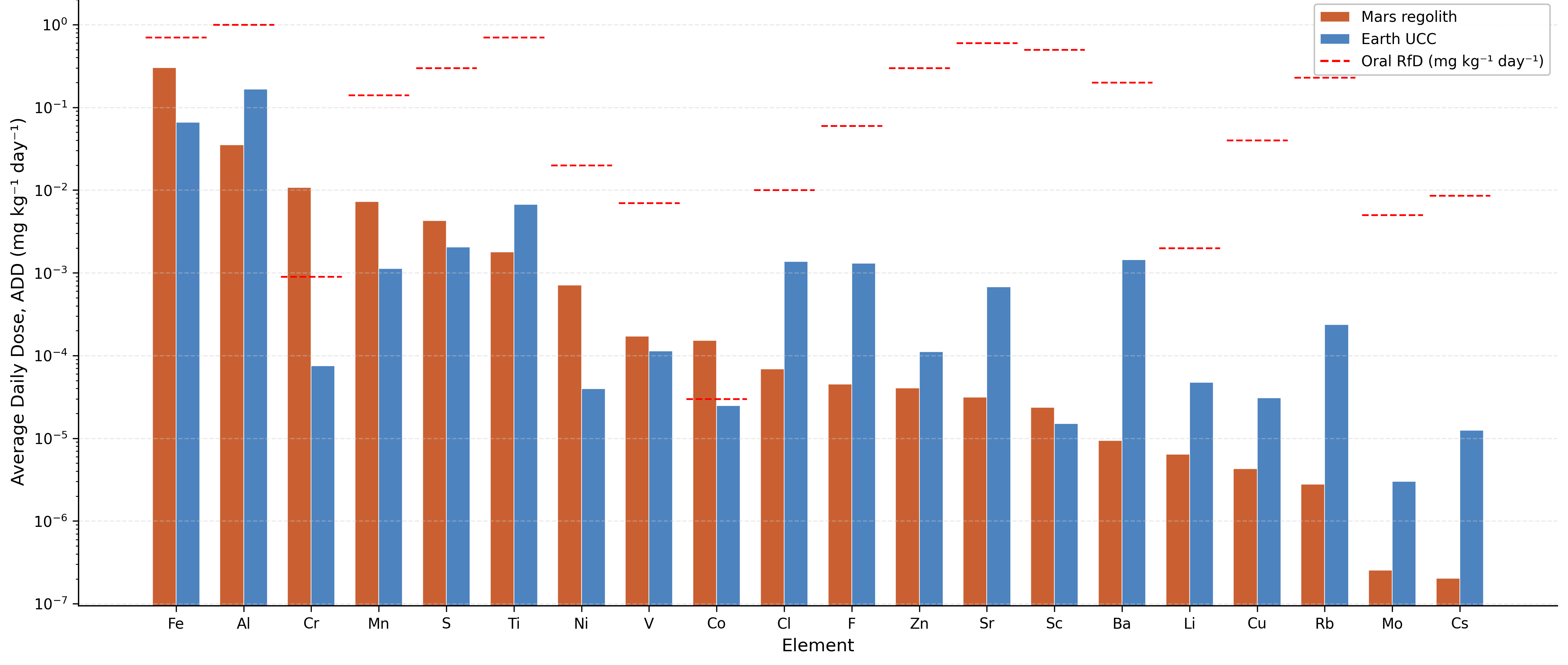}
\end{center}
\caption{Enrichment factor for martian elements normalized to upper continental crust of Earth (Al).\label{fig3}}
\end{figure}

Figure~\ref{fig3} presents the total \acf{ADD}, summed across all three exposure pathways (ingestion, inhalation, and dermal absorption). Iron (Fe) dominates the \ac{ADD} spectrum for Mars by a wide margin, with \ac{ADD}(Mars,Fe)~=~$3.0 \times 10^{-1}$ mg kg$^{-1}$ day$^{-1}$, a direct consequence of its exceptionally high abundance in martian basaltic regolith. Aluminium ranks second at \ac{ADD}(Mars,Al)~=~$3.5 \times 10^{-2}$ mg kg$^{-1}$ day$^{-1}$, and chromium third at \ac{ADD}(Mars,Cr)~=~$1.1 \times 10^{-2}$ mg kg$^{-1}$ day$^{-1}$. Manganese (\ac{ADD}~=~$7.3 \times 10^{-3}$), sulphur (\ac{ADD}~=~$4.3 \times 10^{-3}$), and titanium (\ac{ADD}~=~$1.8 \times 10^{-3}$) follow in descending order. Across all elements, ingestion accounts for 88–92\% of the total \ac{ADD}, inhalation contributes 7–11\%, and dermal absorption remains below 1\% under suit-protected \ac{EVA} conditions. This pathway dominance of ingestion driven by the 100~mg day$^{-1}$ regolith intake assumption has important practical implications: airlock decontamination and post-\ac{EVA} hygiene protocols targeting hand-to-mouth contact pathways are as important as atmospheric filtration in controlling total astronaut dose.

Comparison of Mars \ac{ADD} with Earth \ac{UCC} \ac{ADD} reveals two compositionally distinct groups. Mars \ac{ADD} substantially exceeds \ac{UCC} \ac{ADD} for the mafic-compatible and siderophile elements: Cr exhibits the largest variation, followed by Mn, Fe, Co, Sc, and V. These enrichments directly reflect the absence of crustal differentiation on Mars unlike Earth, which has concentrated compatible elements (Cr, Ni, Co) in the lower crust and mantle through billions of years of arc magmatism and crustal recycling, the martian upper crust retains near-primitive mantle abundances of these elements globally distributed at the surface.

Conversely, Earth \ac{UCC} \ac{ADD} substantially exceeds Mars \ac{ADD} for the 
\acf{LILE} and highly incompatible trace elements that are preferentially concentrated in felsic, differentiated crusts: These include Rb, Ba, K, Cs, Pb, Th and U. These elements are strongly depleted in Mars regolith relative to \ac{UCC} because Mars lacks a granitic upper crust the geological reservoir on Earth that concentrates incompatible elements. The practical consequence is that the list of elements responsible for heavy-metal contamination risk in terrestrial industrial and mining environments (As, Pb, Cd, Hg, Tl) are largely absent as quantitatively significant \ac{ADD} contributors in the martian regolith, with all five elements registering Mars \ac{ADD} values one to three orders of magnitude below their \ac{UCC} counterparts. The dominant geochemical dose burden on Mars is from Fe, Cr, Mn, Co and therefore geologically unique to the martian environment and has no direct terrestrial analogue in terms of combined source-term magnitude and global spatial extent.

\subsection{Non-Carcinogenic Hazard Index}
\subsubsection{HI Across Four Filtration Scenarios}
\begin{figure}[htbp]
\begin{center}
\includegraphics[width=1.0\textwidth]{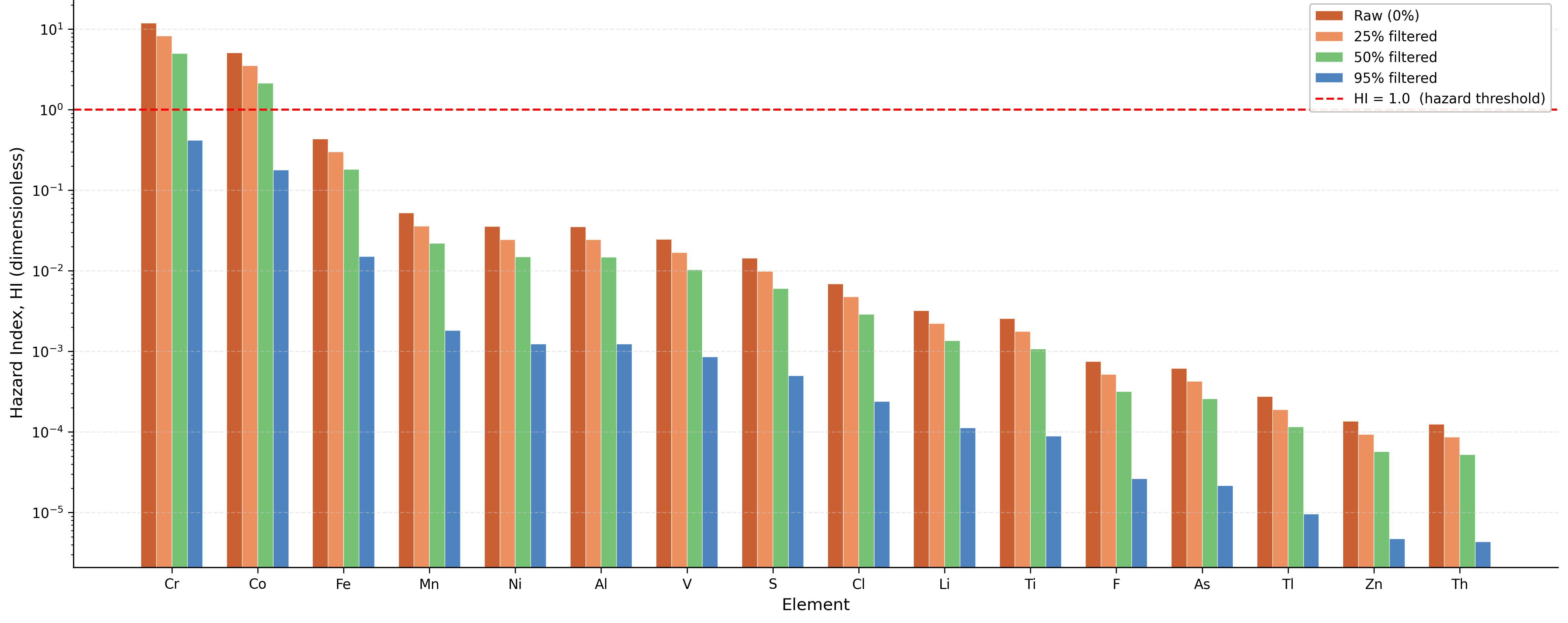}
\end{center}
\caption{Non carcinogenic health hazards (Hazard Index) with detailed filtration scenarios.\label{fig4}}
\end{figure}

Hazard index values are calculated and plotted in Fig.~\ref{fig4}. These values present the non-carcinogenic risks for elements with \ac{EPA} \ac{IRIS} chronic oral \ac{RfD} values modelled across four filtration scenarios: Raw (0\%), 25\%, 50\%, and 95\%. The regulatory threshold of \ac{HI}~=~1, above which non-carcinogenic health effects are considered possible. In the unfiltered (raw) scenario, only chromium (\ac{HI}~=~11.90) and cobalt (\ac{HI}~=~5.08) exceed \ac{HI}~=~1. All other elements remain below \ac{HI}~=~1 in the raw scenario, with the highest third-ranked element being iron (\ac{HI}~=~0.43). This is a notable result: despite Mars being heavily enriched in numerous elements relative to Earth \ac{UCC}, the combination of absolute concentration, oral \ac{RfD}, and mission exposure parameters produces extremely high-risk non-cancer hazard for only two elements. At 25\% filtration, Cr \ac{HI} is reduced to 8.22 and Co \ac{HI} to 3.51, both remain substantially above the threshold. At 50\% filtration, Cr~=~5.01, Co~=~2.14; still hazardous. Only at 95\% filtration do both elements fall below the threshold: Cr~=~0.42, Co~=~0.18. This confirms that no filtration scenario below 95\% is sufficient to bring the hazard indices of both priority elements below the regulatory threshold simultaneously. The progression from raw to 95\% filtered reveals the non-linearity of dose reduction: the first 50\% reduction in filtration yields a reduction in Cr \ac{HI} from 11.90 to 5.01, while the step from 50\% to 95\% filtration yields a further reduction from 5.01 to 0.42. This non-linear response arises because the dose-filtration relationship is linear, but the \ac{HI} threshold crossing is a point condition, making the upper filtration range disproportionately impactful for regulatory compliance.

\begin{figure}[htbp]
\begin{center}
\includegraphics[width=1.0\textwidth]{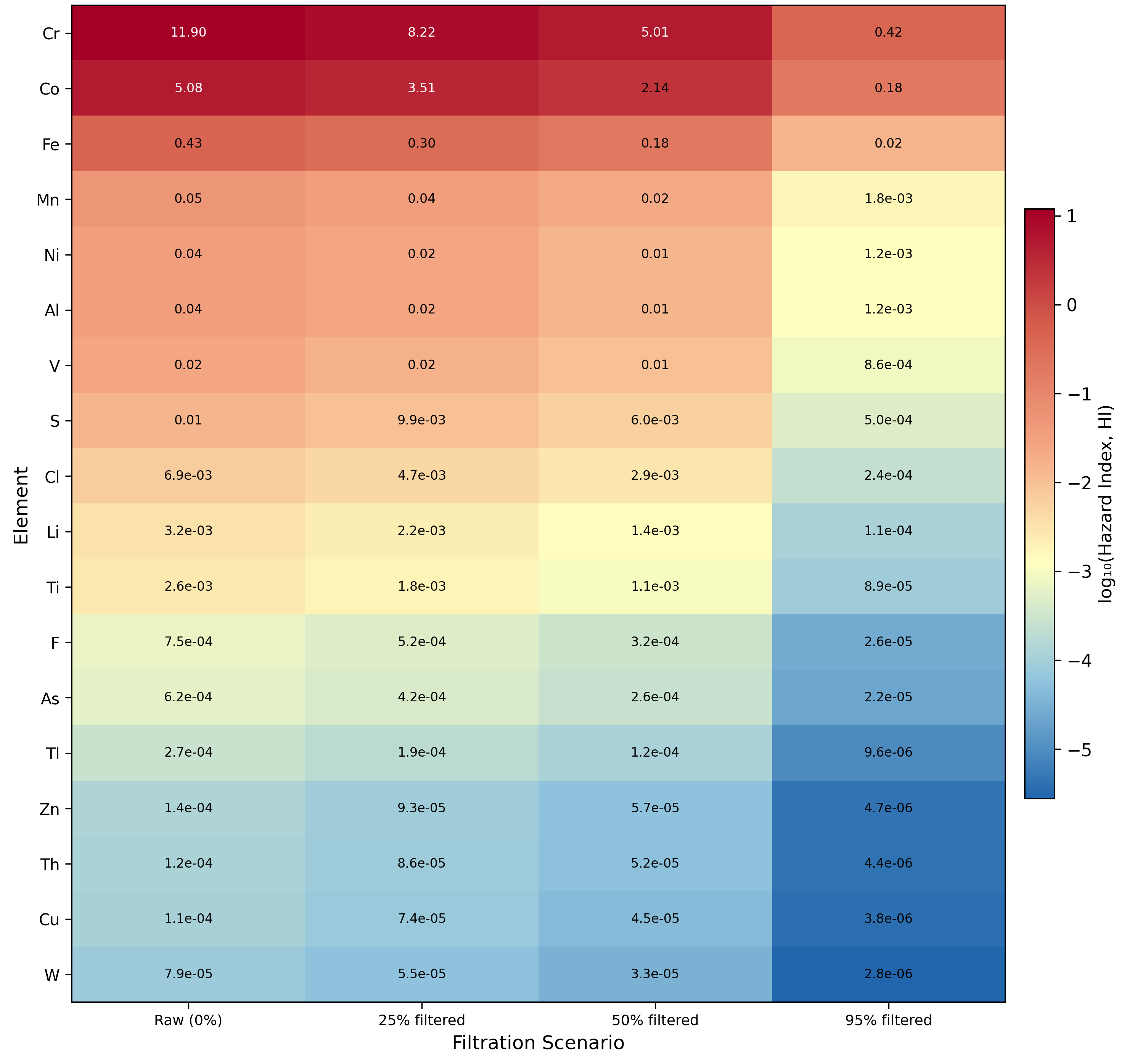}
\end{center}
\caption{Detailed \acf{HI} Heatmap.\label{fig5}}
\end{figure}

The Fig.~\ref{fig5} heatmap displays a comprehensive \acf{HI} matrix for 17 elements across all four filtration scenarios on a $\log_{10}$ colour scale, with numerical \ac{HI} values annotated in each cell. The heatmap reveals a clear tri-level structure in the dataset. Tier I elements (Cr, Co) are distinguished by warm red colouring in the Raw and 25\% columns (\ac{HI} $>$ 1), transitioning through orange at 50\% to yellow at 95\%. Tier II elements (Fe: \ac{HI}~=~0.43 raw; Mn: \ac{HI}~=~0.05 raw; Ni: \ac{HI}~=~0.04 raw; Al: \ac{HI}~=~0.04 raw; V: \ac{HI}~=~0.02 raw) occupy the \ac{HI} range 0.001–1.0 in the raw scenario, shown in transitional yellow-green tones. Tier III elements (S, Cl, Li, Ti, F, As, Tl, Zn, Th, Cu, W) have raw \ac{HI} values in the range $10^{-4}$ to $10^{-2}$, shown in cold blue. The heatmap also confirms the systematic proportional reduction of \ac{HI} with filtration across all elements: each 95\% filtration step reduces all \ac{HI} values by approximately one order of magnitude (factor $\sim 20 \times$), which is consistent with the linear dose-filtration model. An important observation from the heatmap is that elements in Tier II (Fe, Mn, Ni, Al, V) could approach or breach \ac{HI}~=~1 under scenarios not modelled here, such as elevated ingestion rates (e.g., 200 mg/day for spacesuit leakage events) or longer mission durations (e.g., 3-year conjunction-class missions), warranting secondary monitoring protocols even for these nominally sub-threshold elements.

\subsubsection{Mars vs.\ Earth UCC HI Comparison} 

\begin{figure}[htbp]
\begin{center}
\includegraphics[width=1.0\textwidth]{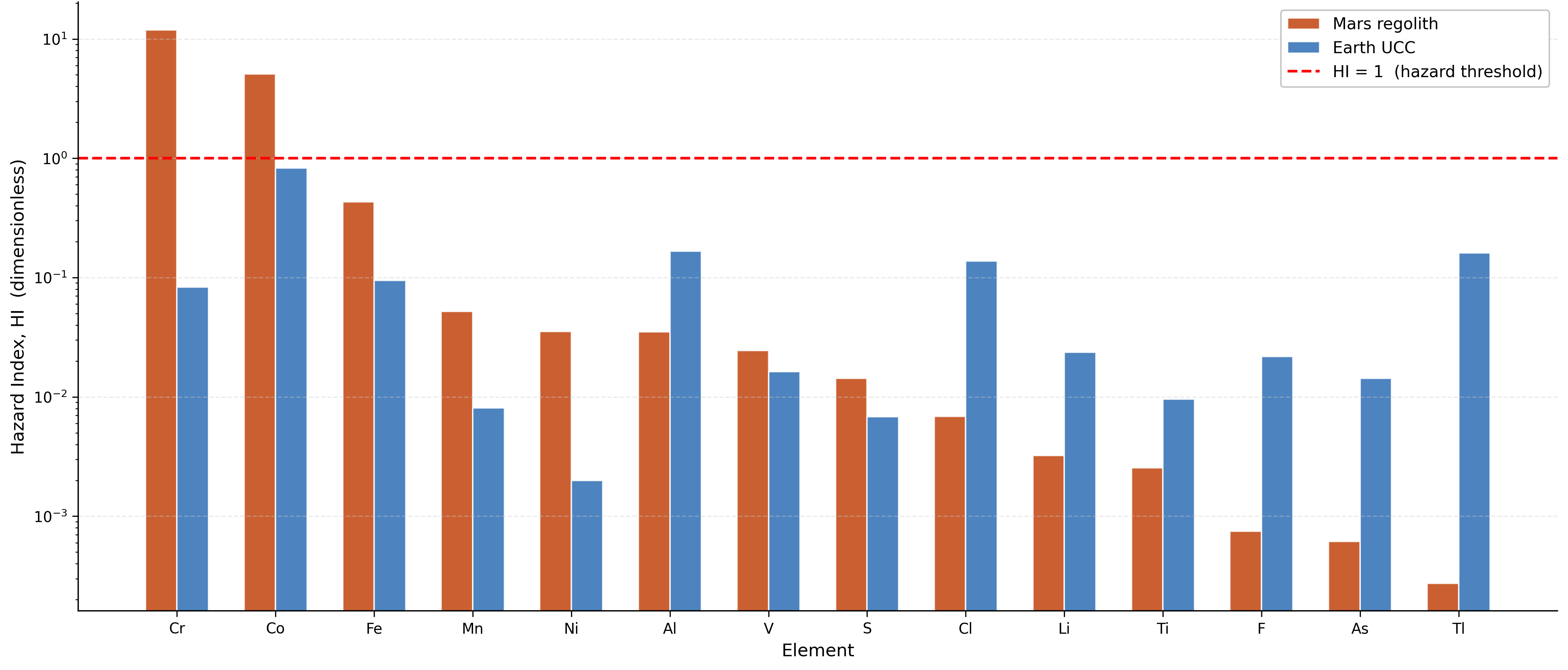}
\end{center}
\caption{Hazard Index Comparison plot for martian regolith and \ac{UCC}.\label{fig6}}
\end{figure}

To compare the non-carcinogenic health hazards presented by martian regolith with the Upper continental crust, The modelled \ac{HI} values are given in Fig.~\ref{fig6}, which directly compares the non-carcinogenic \ac{HI} of Mars regolith to that of Earth \ac{UCC} at equivalent exposure parameters, for the 14 highest-ranking elements by Mars \ac{HI}. The comparison underscores the unique nature of the martian hazard profile. For chromium, the Mars \ac{HI} (11.90) exceeds the equivalent \ac{UCC}-based \ac{HI} (approximately 0.08) by a factor of approximately 149. This confirms that the Cr hazard is driven entirely by the extraordinary geochemical enrichment of Cr in the martian mantle rather than by any methodological peculiarity. For cobalt, Mars \ac{HI} (5.08) exceeds \ac{UCC} \ac{HI} (approximately 0.79) by a factor of 6.4. The near-unity ratio of \ac{HI} excess to concentration ratio for both Cr and Co validate the dose-response model and confirms the internal consistency of the assessment.

A counterintuitive result visible in Fig.~\ref{fig6} is that Earth \ac{UCC} \ac{HI} exceeds Mars \ac{HI} for aluminium, chlorine, and thallium. For Al, this reflects the higher \ac{UCC} concentration combined with a moderately restrictive \ac{RfD} of 1.0 mg kg$^{-1}$ day$^{-1}$. For chlorine, the \ac{UCC} concentration is 640 ppm vs.\ 32 ppm for Mars, giving \ac{UCC} Cl \ac{HI} approximately 20$\times$ higher than Mars. For thallium (Tl), despite Mars being strongly depleted, the very low TI \ac{RfD} produces non-trivial HQ values at the \ac{UCC} level. None of the Earth \ac{UCC} \ac{HI} values shown in Figure 7 exceed the \ac{HI}~=~1 threshold, confirming that standard terrestrial regolith presents no regulatory non-cancer risk in this exposure model, and that the hazards identified for Mars are genuinely Mars-specific.

\subsubsection{Cumulative Hazard Index}

\begin{figure}[htbp]
\begin{center}
\includegraphics[width=1.0\textwidth]{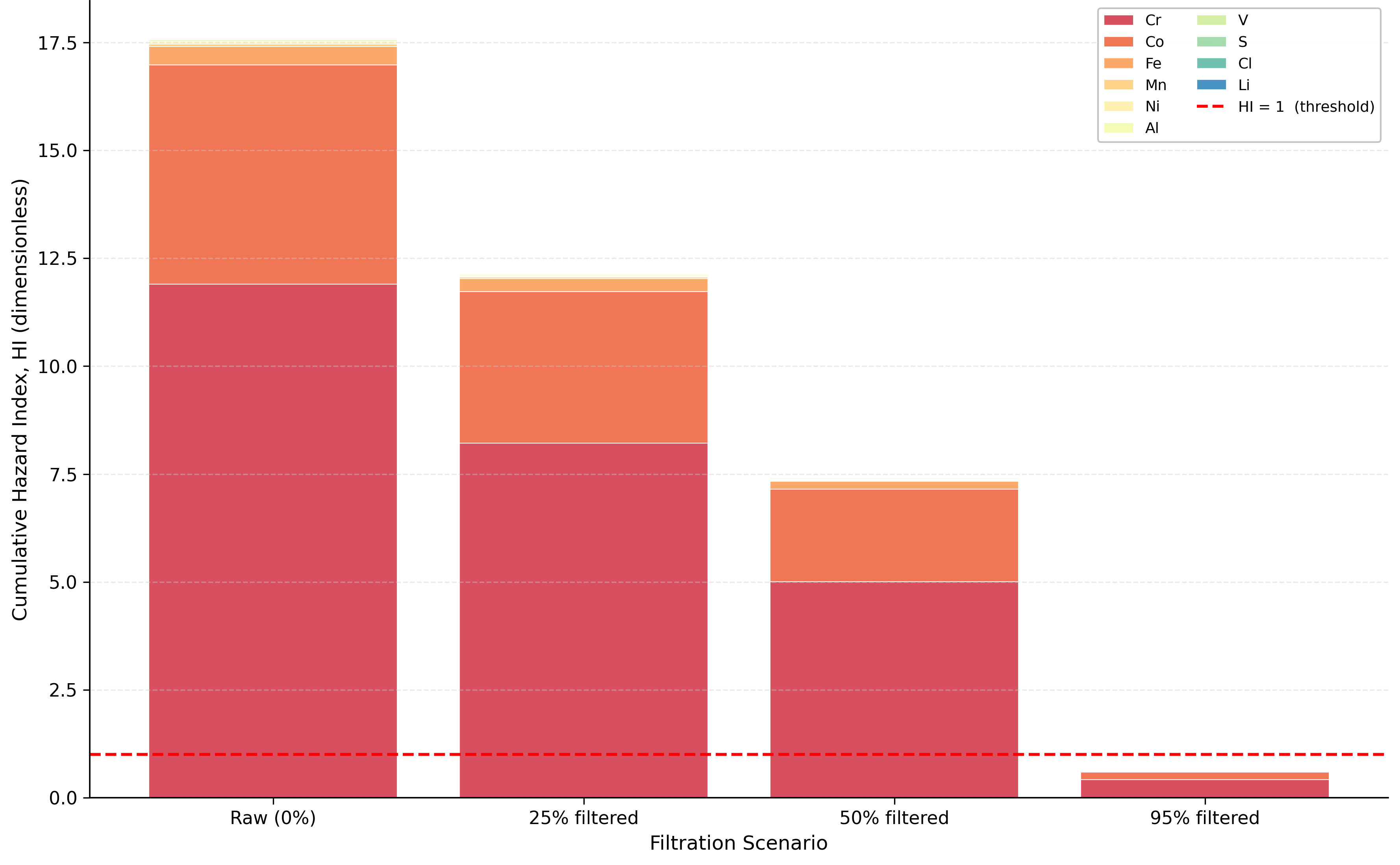}
\end{center}
\caption{Cumulative (Total) hazard index for martian regolith with various filtration levels.\label{fig7}}
\end{figure}

The cumulative hazard index values for martian regolith are plotted in Fig.~\ref{fig7}, showing the elemental contributions to total system-level risk across all four filtration scenarios. In the raw scenario, the cumulative \ac{HI} reaches approximately 17.5, dominated by Cr (contributing approximately 11.9/17.5~=~68\% of the total) and Co (approximately 5.1/17.5~=~29\%). The combined Cr + Co fraction exceeds 97\% of total cumulative \ac{HI}, confirming that these two elements are the near-complete determinants of the non-cancer hazard. At 25\% filtration, the cumulative \ac{HI} falls to approximately 12.0; at 50\% to approximately 7.3; and at 95\% to approximately 0.62. The crossing of the cumulative \ac{HI}~=~1 threshold occurs between 50\% and 95\% filtration, confirming that the 95\% filtration scenario is the minimum standard filter capable of reducing the aggregate non-cancer burden below the regulatory threshold. The stacked chart additionally confirms that below Cr and Co, the contributions of Fe, Mn, Ni, Al, and other elements to total risk are minor ($\leq$ 3\% combined), simplifying the risk management target to two primary elements.

\subsubsection{Minimum Filtration Efficiency Required} 

The minimum regolith filtration efficiency required to reduce HI below 1.0 for the two elements exceeding this threshold in the unfiltered scenario is calculated and the values are given in Table~\ref{tab1}. For chromium, a minimum filtration efficiency of 91.6\% is required. For cobalt, a minimum filtration efficiency of 74.3\% is required. The result that Cr requires $>$91.6\% filtration and Co requires $>$74.3\% filtration has a fundamental engineering implication: a single-stage 95\% \ac{HEPA} filter satisfies the non-cancer threshold for both elements simultaneously, provided that surface decontamination (addressing the dominant ingestion pathway) achieves equivalent efficiency. High-efficiency particulate air (\ac{HEPA}) filters per the \ac{NIOSH} standard achieve $\geq$99.97\% efficiency for particles $\geq$0.3~$\mu$m \citep{anjum2024m}, which technically exceeds the minimum required for both Cr and Co. However, nano-scale martian dust ($\geq$0.1~$\mu$m), including nanophase iron oxides identified by \citet{morris2006}, may penetrate standard \ac{HEPA} filters, suggesting that combined \ac{HEPA}/electrostatic or \ac{HEPA}/activated-carbon systems should be specified.

\begin{table}
\begin{tabular}{|c|c|c|}
\hline
Element	& Hazard Index (Unfiltered) &	Minimum Required Filtration \% to get \ac{HI}$<$1) \\
\hline
Cr  &	11.9  &   91.6 \\
Co	&    5.1 &   74.3 \\
\hline
\end{tabular}
\caption{Minimum Filtration required for Cr and Co to reduce the risk of Non-Carcinogenic Health Hazards Threshold.\label{tab1}}
\end{table}

\subsection{Carcinogenic Risk Assessment}
\subsubsection{ILCR: Mars vs.\ Earth \ac{UCC}}
\begin{figure}[htbp]
\begin{center}
\includegraphics[width=1.0\textwidth]{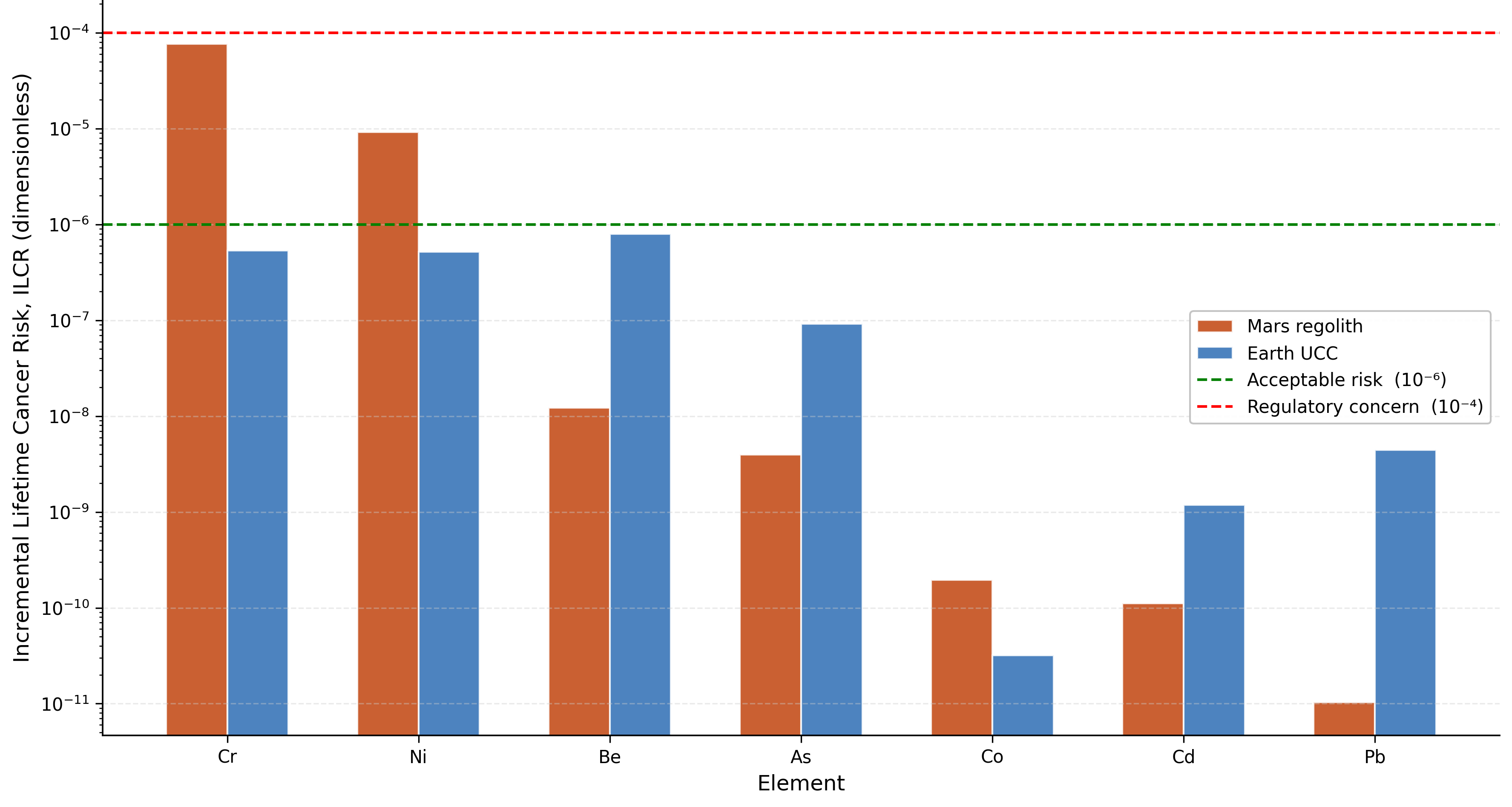}
\end{center}
\caption{\acf{ILCR} values plotted for Mars and Earth's \ac{UCC}.\label{fig8}}
\end{figure}

A long-term stay on the martian surface could be harmful for an astronaut's health and may pose carcinogenic risks from martian surface radiation and exposure to toxic metals from the martian regolith. The \acf{ILCR} is calculated for the toxic metals exposure and the values are plotted in Fig.~\ref{fig8} using \ac{EPA} \ac{IRIS} \acf{CSF} \citep{epairis}, comparing martian regolith to the \ac{UCC} of Earth in the unfiltered scenario. Two elements exceed the acceptable cancer risk threshold of $10^{-6}$ from martian regolith exposure: chromium (\ac{ILCR}~=~$5.8 \times 10^{-5}$) and nickel (\ac{ILCR}~=~$9.2 \times 10^{-6}$). Chromium's \ac{ILCR} of $5.8 \times 10^{-5}$ is noteworthy: it is above the $10^{-6}$ acceptable threshold and approaches the $10^{-4}$ regulatory concern level. The EPA oral CSF for Cr(VI) of 0.5 (mg kg$^{-1}$ day$^{-1}$)$^{-1}$, combined with the ADD of $1.1 \times 10^{-2}$ mg kg$^{-1}$ day$^{-1}$, produces this elevated \ac{ILCR}. The analysis assumes that all martian Cr is present as Cr(VI), the more toxic hexavalent form; if a proportion exists as Cr(III), the actual \ac{ILCR} would be lower, but the highly oxidising surface conditions on Mars, reported by \citet{tuff2013} support a significant Cr(VI) fraction. Nickel's \ac{ILCR} of $9.2 \times 10^{-6}$ places it just above the $10^{-6}$ acceptable threshold. This is driven by the elevated martian Ni and the relatively high \ac{CSF} for nickel of 0.91 (mg kg$^{-1}$ day$^{-1}$)$^{-1}$. Importantly, Fig.~\ref{fig8} (Cancer Risk) shows that for arsenic, Earth \ac{UCC} actually produces a higher \ac{ILCR} (approximately $1.1 \times 10^{-7}$) than Mars (approximately $4.0 \times 10^{-9}$), directly reflecting the strong depletion of As on Mars. Similarly, for Pb and Cd, Earth \ac{UCC} produces higher \ac{ILCR}s, countering a common assumption that Mars would be more carcinogenic across the board. For Be, it exhibits an \ac{ILCR} of approximately $1.0 \times 10^{-8}$ from Mars, far below the $10^{-6}$ threshold, despite Be's exceptionally high \ac{CSF} of 8.4 (mg kg$^{-1}$ day$^{-1}$)$^{-1}$. This is because martian Be is strongly depleted as compared to \ac{UCC}, the depletion factor more than offsets the high \ac{CSF}. 

\subsubsection{ILCR Across Filtration Scenarios}
\begin{figure}[htbp]
\begin{center}
\includegraphics[width=1.0\textwidth]{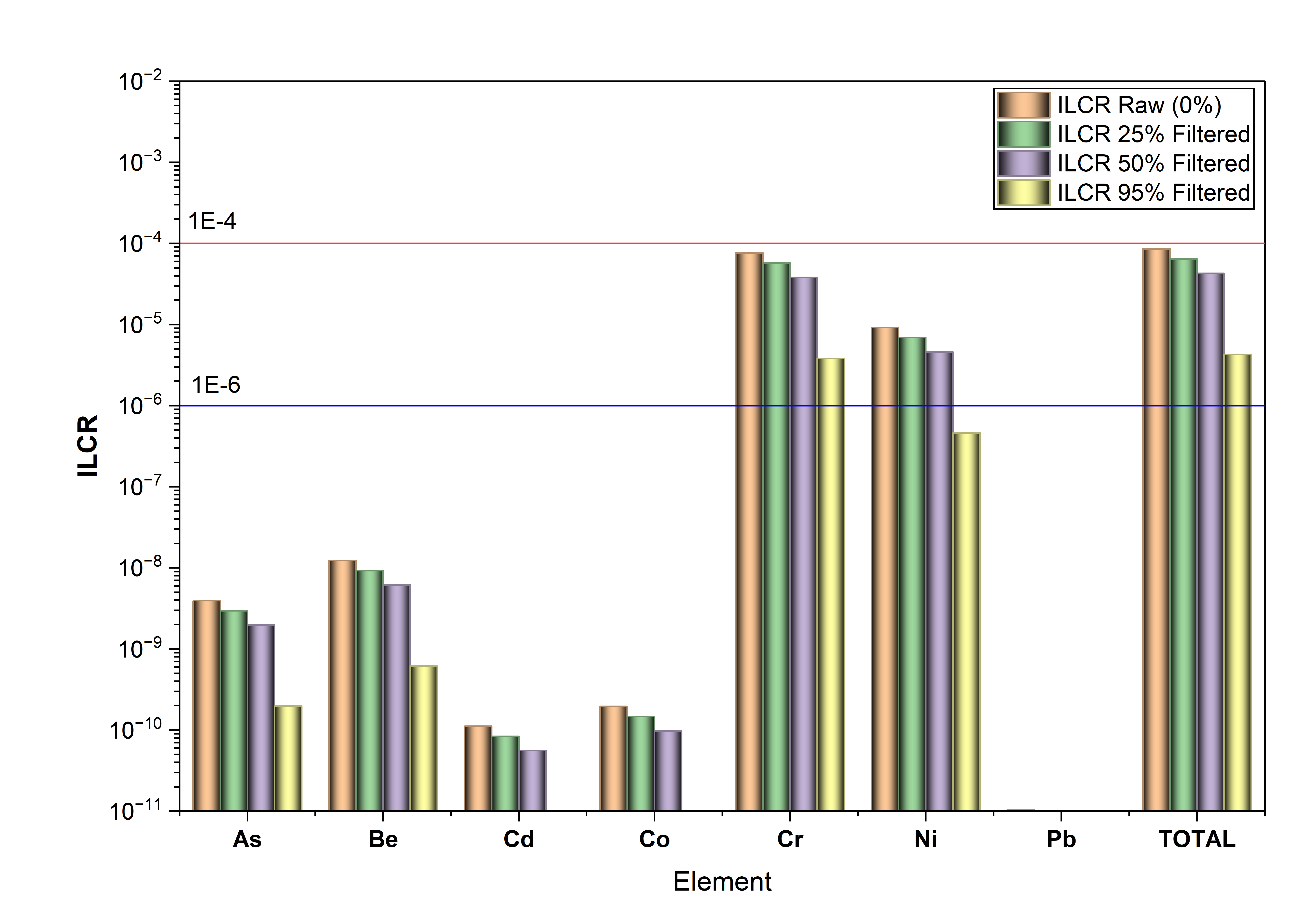}
\end{center}
\caption{ILCR values calculated and graphed with various filtration ranges.\label{fig9}}
\end{figure}

The total \acf{ILCR} across various filtration scenarios is plotted in Fig.~\ref{fig9}. The pattern confirms that Cr dominates at all filtration levels: at 0\% (raw), \ac{ILCR}~=~$5.8 \times 10^{-5}$; at 25\%, \ac{ILCR}~=~$4.3 \times 10^{-5}$; at 50\%, \ac{ILCR}~=~$2.9 \times 10^{-5}$; and at 95\%, \ac{ILCR}~=~$2.9 \times 10^{-6}$. Critically, even at 95\% filtration, the Cr \ac{ILCR} remains almost $\sim$3 times above the acceptable $10^{-6}$ threshold. Nickel follows a similar trend: raw \ac{ILCR}~=~$9.2 \times 10^{-6}$; 25\%: $6.8 \times 10^{-6}$; 50\%: $4.6 \times 10^{-6}$; 95\%: $4.6 \times 10^{-7}$. At 95\% filtration, Ni \ac{ILCR} crosses below the $10^{-6}$ threshold to $4.6 \times 10^{-7}$, establishing 95\% as the minimum standard for Ni cancer risk acceptability and long term astronaut safety. All other elements (As, Be, Cd, Co, Pb) remain below $10^{-6}$ at all filtration levels, including the unfiltered scenario. The residual Cr carcinogenic risk at 95\% filtration (\ac{ILCR}~=~$2.9 \times 10^{-6}$) represents a persistent, non-eliminable cancer risk under any practical filtration scenario. Total elimination of Cr cancer risk to below $10^{-6}$ would theoretically require $>$99\% filtration efficiency, which is technically achievable but not practically guaranteed in Mars surface operations due to filter loading, bypass leakage, and the nano-scale particle fraction. This finding is consistent with the conclusion of \citet{wang2025} that medical counter measures, specifically Vitamin C supplementation to reduce Cr(VI) to the less carcinogenic Cr(III) form, should be considered as a complementary risk reduction strategy alongside engineering controls. When we look at the cumulative \ac{ILCR}, it is concerning that the total risk remains above the US \ac{EPA} threshold even after 95\% filtration, implying that for long missions on Mars, the filtration efficiency should be greater than 95\% to reduce the cancer risk for astronauts via martian regolith ingestion, inhalation, and skin contact.

\subsection{Monte Carlo Probabilistic Risk Analysis}
\subsubsection{Probabilistic Hazard Index Distributions}

\begin{figure}[htbp]
\begin{center}
\includegraphics[width=1.0\textwidth]{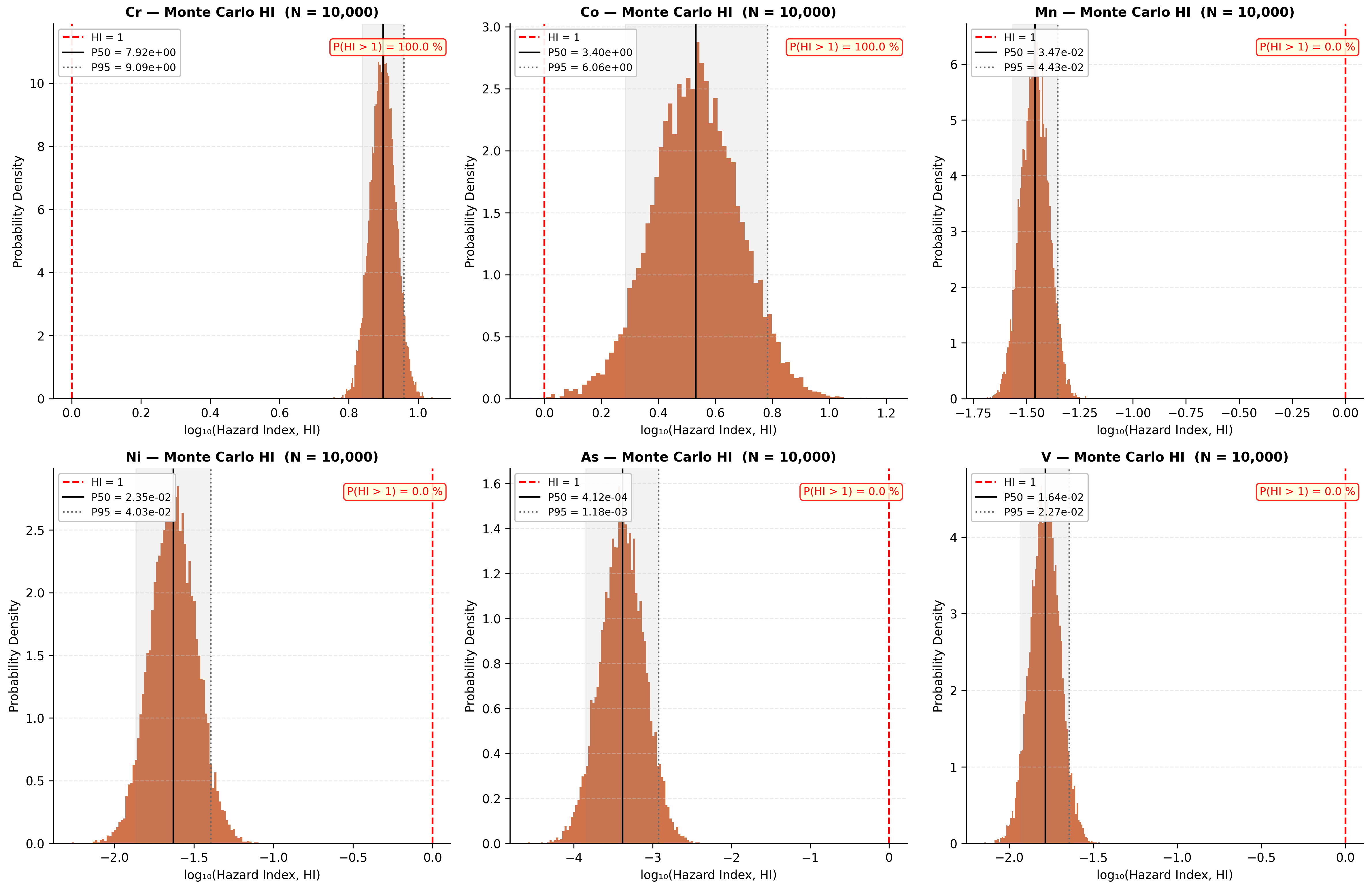}
\end{center}
\caption{Probabilistic Threshold for non-carcinogenic health hazards (HI).\label{fig10}}
\end{figure}

To quantify the robustness of the major risk posing elements for non-carcinogenic health hazards, probabilistic health hazards were quantified. Fig.~\ref{fig10} (Monte Carlo) presents the probabilistic Hazard Index distributions from N~=~10,000 Monte Carlo simulations for six elements (Cr, Co, Mn, Ni, As, V), each with elemental concentrations sampled from log-normal distributions parameterised by the Taylor values and specific \ac{CV} which were parametrized by Taylor, derived from \ac{GRS} orbital measurement uncertainties and inter-laboratory meteorite data. For chromium (\ac{CV}~=~0.084), the tight log-normal distribution produces a Monte Carlo P50~=~7.92 (median \ac{HI}) and P95~=~9.09, with P(\ac{HI}~$>$~1)~=~100.0\%. The narrow distribution reflects \ac{GRS}'s relatively low measurement uncertainty for Cr at the global scale. The entire probability mass is concentrated between $\log_{10}$(\ac{HI})~=~0.3 and 1.0, far above the \ac{HI}~=~1 threshold ($\log_{10}$~=~0). This result is among the most robust in the analysis: even at the extreme lower end of the geochemical uncertainty distribution (P5~$\approx$~6.0), Cr \ac{HI} remains approximately 6$\times$ above the threshold. For cobalt (\ac{CV}~=~0.35), the broader distribution yields P50~=~3.40, P95~=~6.06, and P(\ac{HI}~$>$~1)~=~100.0\%. The wider distribution compared to Cr reflects greater meteorite inter-laboratory variability in Co determinations. The distribution is well-separated from the \ac{HI}~=~1 threshold, with essentially no probability mass below it. Manganese shows P50~=~0.0347, P95~=~0.0443, and P(\ac{HI}~$>$~1)~=~0.0\%, confirming it as robustly sub-threshold across its full uncertainty distribution. Nickel (\ac{CV}~=~0.33) yields P50~=~0.0235, P95~=~0.0403, and P(\ac{HI}~$>$~1)~=~0.0\%, despite Ni's extremely high EF and carcinogenic \ac{ILCR} exceeding 10$^{-6}$. This apparent paradox reflects the relatively permissive Ni oral \ac{RfD} of 0.02 mg kg$^{-1}$ day$^{-1}$ \citep{hadjiliadis2012}, which produces \ac{HQ} values well below 1 even at martian concentrations. Arsenic (\ac{CV}~=~0.64, P50~=~$4.12 \times 10^{-4}$, P(\ac{HI}~$>$~1)~=~0.0\%) and vanadium (\ac{CV}~=~0.20, P50~=~$1.64 \times 10^{-2}$, P(\ac{HI}~$>$~1)~=~0.0\%) similarly show no probability of exceeding the non-cancer threshold.

The Monte Carlo results thus confirm with high statistical certainty that the deterministic single-point analysis is not misleading: the conclusion that Cr and Co are the only elements exceeding \ac{HI}~=~1 is robust to the full range of plausible geochemical uncertainty. This is a critical finding for mission planning and Human safety.

\subsubsection{Probabilistic Cancer Risk Distributions }

\begin{figure}[htbp]
\begin{center}
\includegraphics[width=1.0\textwidth]{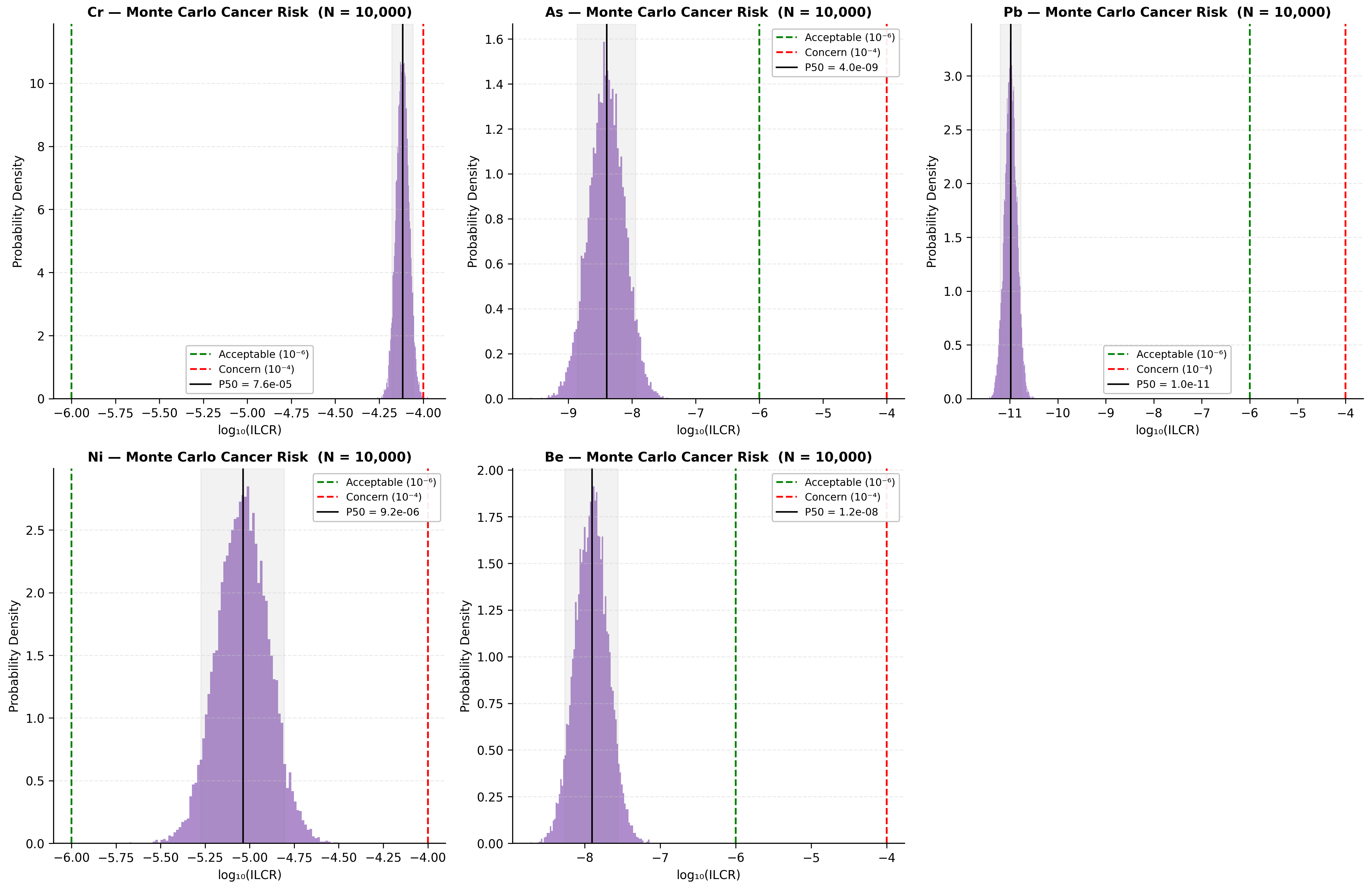}
\end{center}
\caption{Monte carlo probabilistic cancer occurrence risk threshold.\label{fig11}}
\end{figure}

Similarly, the probabilistic threshold value for \ac{ILCR} were quantified. These values are given in Fig.~\ref{fig11}. This figure presents Monte Carlo based probabilistic \ac{ILCR} distributions for five carcinogenic elements. The Cr distribution (\ac{CV}~=~0.084) produces P50~=~$7.6 × 10^{-5}$, with 100\% of simulations above the $10^{-6}$ threshold (visible on the $\log_{10}$ scale where the entire distribution lies well above $10-6$). The P5 and P95 bound approximately 95\% of the probability mass within the range $4 \times 10^{-5}$ to $1.0 \times 10^{-4}$, with the upper tail approaching the $10^{-4}$ regulatory concern level. Nickel (\ac{CV}~=~0.33) produces P50~=~$9.2 \times 10^{-6}$, with essentially 100\% of simulations above $10^{-6}$. The distribution is broad relative to Cr owing to Ni's higher \ac{CV}, spanning approximately two orders of magnitude (P5~=~$2 \times 10^{-6}$ to P95~=~$3 \times 10^{-5}$), yet almost entirely above the threshold. The Arsenic (\ac{CV}~=~0.64, P50~=~$4.0 \times 10^{-9}$), lead (\ac{CV}~=~0.30, P50~=~$1.0 \times 10^{-11}$), and beryllium (\ac{CV}~=~0.50, P50~=~$1.0 \times 10^{-8}$) all remain well below $10^{-6}$ with P(\ac{ILCR}~$>$~$10^{-6}$)~$\approx$~0\% even considering their high concentration uncertainty.

The probabilistic analysis therefore confirms that the carcinogenic risk conclusions are robust: Cr and Ni are the only elements with confirmed (P~=~100\% and ~100\% respectively) carcinogenic risk exceeding the $10^{-6}$ threshold, while As, Pb, and Be do not constitute carcinogenic concerns under the modelled martian surface conditions.

\section{Discussion}\label{sec:discussion}
\subsection{Chromium as the Dominant Health Risk: Geochemical and Toxicological Basis}

The convergence of all risk metrics \ac{EnFac} (673), \ac{ERF} (716), \ac{HI} (11.90 raw), \ac{ILCR} ($5.8 \times 10^{-5}$), and Monte Carlo probability of hazard threshold P(\ac{HI}~$>$~1)~=~100\% unanimously identifies chromium as the most critical geochemical health hazard in martian regolith. This finding is mechanistically grounded in the fundamental differences in planetary differentiation between Earth and Mars. On Earth, more than 4.5 billion years of subduction, arc magmatism, and crustal recycling have progressively concentrated incompatible elements (K, Th, U, \ac{REE}) in the continental crust while sequestering compatible and siderophile elements (Cr, Co, Ni, Mn, Fe) into the lower crust, subcontinental lithosphere, and mantle. Mars, lacking plate tectonics and operating with a far lower degree of magmatic differentiation, has preserved a near-chondritic Cr abundance (4,990~ppm) in its accessible crust, compared to only 35~ppm in Earth's \ac{UCC} \citep{treiman2016, gaschnig2016}.

The toxicological significance of this geochemical distinction is compounded by the oxidation state of the martian surface. The work of \citet{tuff2013} demonstrated, from the oxygen fugacity recorded in martian shergottite meteorites, that the martian upper mantle was oxidised at approximately \ac{FMQ}+1 to \ac{FMQ}+3 significantly more oxidising than Earth's mantle at approximately \ac{FMQ} to \ac{FMQ}+0.5. Under these conditions, a high portion of crustal Cr is expected to be present as Cr(VI), the hexavalent state classified by the \ac{IARC} as a Group~1 human carcinogen \citep{epairis, wang2017}. The proportion of Cr(VI) vs.\ Cr(III) at the martian surface is not yet directly measured and is a critical unknown for quantitative risk assessment. This analysis uses the Cr(VI) \ac{CSF} as a conservative upper-bound assumption; if future in situ spectroscopic measurements (e.g., \ac{XANES}) reveal a predominantly Cr(III) surface, the carcinogenic \ac{ILCR} would be reduced significantly, but the evidence for highly oxidizing martian environment makes this less likely.

From a clinical point of view, chronic exposure to Cr(VI) in occupational settings results in pulmonary sensitivity, nasal septum perforation, and a dose-dependent increase in the risk of lung cancer, as confirmed by epidemiological cohort studies of chromate production workers \citep{junaid2016}. The colon and stomach lining are the primary target tissues for localized oxidative DNA damage caused by gastrointestinal Cr(VI) absorption, which, accounts for ~90\% of total \ac{ADD} in the martian context. Chronic oral Cr(VI) exposure leads to non-carcinogenic systemic effects, such as hepatotoxicity, renal tubular injury, and haematological effects, including methaemoglobinaemia at high doses \citep{katsas2024}. The martian \ac{ADD} of $1.1 \times 10^{-2}$~mg~kg$^{-1}$~day$^{-1}$ exceeds the \ac{EPA} chronic oral Reference Dose for Cr(VI) of $9 \times 10^{-4}$~mg~kg$^{-1}$~day$^{-1}$ by a factor of approximately 12 in the unfiltered scenario, resulting in \ac{HI}~=~11.90.

From martian Regolith, Cr(VI) presents a dual clinical challenge: it must be managed as both a non-carcinogenic systemic toxicant and a carcinogen for which no safe threshold exposure exists under the linear no-threshold model used in \ac{EPA} risk assessment. The primary intervention stratum is represented by engineering controls, specifically \ac{HEPA}/\ac{ULPA} filtration at $\geq$95\% efficiency and systematic airlock decontamination, which are discussed in the Recommendations section. Nevertheless, the residual Cr \ac{ILCR} of $2.9 \times 10^{-6}$ at 95\% filtration, which is 2.9-fold above the acceptable threshold, requires the implementation of supplementary medical countermeasures.

\citet{wang2025} suggested for prophylactic Vitamin C (ascorbic acid) intake as a medicinal approach to convert Cr(VI) into the significantly less toxic and non-carcinogenic Cr(III) form at the gastrointestinal mucosal surface before systemic absorption occurs. The suggested mechanism utilizes the well-documented in vitro and in vivo reduction of Cr(VI) by ascorbate at physiological pH, resulting in the formation of Cr(III)-ascorbate complexes with reduced bioavailability \citep{xu2005}. A supplementation dosage of 500–1,000~mg daily within the tolerated upper intake threshold set by the Institute of Medicine has been proposed as an effective preventive strategy \citep{johnston1999}. Vitamin C supplementation in microgravity presents a secondary risk of heightened urinary oxalate excretion, thereby increasing the likelihood of calcium oxalate kidney stone formation, which is already elevated in astronauts due to hypercalciuria linked to bone demineralization \citep{baxmann2003, patel2020}. This pharmaceutical intervention necessitates a detailed clinical pharmacology assessment within the comprehensive astronaut health profile prior to its implementation as a conventional countermeasure.

Biological monitoring of urine Cr(VI) as a biomarker for internal exposure is standard practice in chromate-related occupations and should be integrated into routine medical surveillance for Mars missions. It is advised to conduct baseline urine chromium measurements prior to the mission, monitor at three-month intervals during the mission, and do post-mission monitoring for a minimum of 24 months, in accordance with \ac{NIOSH} guidelines for workers exposed to chromium(VI). The occupational exposure limit for Cr(VI) set by \ac{NIOSH} at 0.2~$\mu$g~m$^{-3}$ shall act as the provisional habitat air quality standard for martian surface operations until a Mars-specific occupational exposure limit is developed \citep{cdc2013}. The management implications are correspondingly severe. Achieving \ac{HI}~$<$~1 for Cr necessitates a minimum filtration efficiency of 91.6\%, and reducing \ac{ILCR} below the $10^{-6}$ acceptable threshold necessitates an efficiency of over 99\%. However, this standard cannot be guaranteed under operational martian surface conditions due to filter loading, bypass leakage, and the nano-scale particle fraction ($\leq$~0.1~$\mu$m) as a component of martian dust. 

\subsection{Cobalt as a Previously Overlooked Priority Hazard}

The identification of cobalt as the second most significant non-carcinogenic risk, with a hazard index (\ac{HI}) of 5.08 unfiltered and a probability (P(\ac{HI}~$>$~1)) of 100\% in Monte Carlo simulations, constitutes a notable advancement compared to previous studies. Wang et al. [15], whose review represents the most thorough previous clinical evaluation of the health effects of martian dust, identified Cr, Be, As and Cd as the four key potentially dangerous elements of concern. Cobalt is absent from their priority list, despite possessing a significantly higher deterministic \ac{HI} value than any other non-Cr elements in this dataset. This study discusses the quantitative rationale for elevating Co to co-equal priority with Cr in the health monitoring systems for martian astronauts.

The rationale behind Co's elevated \ac{HI} is not just geochemical enrichment—despite Co being enriched 6.1-fold compared to \ac{UCC} but also a rather remarkably low \ac{EPA} \ac{IRIS} chronic oral Reference Dose of $3.0 \times 10^{-5}$~mg~kg$^{-1}$~day$^{-1}$, one of the the lowest among all elements in the dataset. This \ac{RfD} indicates the established correlation between chronic cobalt exposure and \ac{HMLD}, a progressive and frequently irreversible pulmonary interstitial fibrosis alongside cardiomyopathy, polycythaemia, hypothyroidism, and peripheral neuropathy observed in workers within the established tungsten carbide sector \citep{lison1996cr, lison1996er, linna2004}. The martian Co \ac{ADD} is roughly five times the \ac{RfD}, indicating a significant non-cancer risk. The minimal filtration necessary to reduce Co \ac{HI} below 1.0 is 74.3\%, indicating that a ~75\% efficient filter would sufficiently mitigate the Co non-cancer danger. Meanwhile, the more rigorous Cr criterion of 91.6\% concurrently meets the Co criteria, so appropriately aligning the design standard with the Cr threshold.

The principal clinical issue associated with cobalt in the surface of Mars is pulmonary: hard metal lung disease is characterized by lymphocyte-mediated granulomatous inflammation and progressive fibrosis of the alveolar interstitium, induced by cobalt ions released from inhaled particles that interact with tungsten carbide \citep{lison1996cr} or, in martian regolith, with iron oxide phases that similarly promote the generation of reactive oxygen species \citep{fubini2003}. Cardiac effects, particularly dilated cardiomyopathy linked to cobalt-induced mitochondrial dysfunction, have been documented in occupational environments at blood cobalt concentrations around 300~$\mu$g~L$^{-1}$, significantly exceeding standard environmental exposure levels but potentially pertinent for prolonged habitation on the martian surface \citep{paustenbach2013, jenkinson2021}. Hypothyroidism resulting from prolonged cobalt exposure due to competitive restriction of iodine uptake by the thyroid, a process similar to that of perchlorate, is a new endocrine risk that compounds the known perchlorate hazard present in martian regolith at 0.5–1 weight\% \citep{simonsen2012, paustenbach2013, hecht2009}.

Medical countermeasures for Co exposure are less pharmacologically selective than those for Cr(VI). There is currently no approved chelating agent that has proven effective against chronic low-level cobalt poisoning \citep{paustenbach2013}. N-acetylcysteine has been examined in vitro as an antioxidant that partially mitigates cobalt-induced oxidative stress in lung cells \citep{luczak2013, liu2016}, although clinical data for preventive N-acetylcysteine supplementation in occupational cobalt exposure is scarce. The principal management technique for Co relies on engineering controls, filtration and surface decontamination augmented by biological monitoring of serum and urinary Co as indicators of internal dosage. \citet{linna2004} established reference values for cobalt biomarkers in the hard metal industry, which should be modified for the martian mission context through site-specific exposure modeling. Pulmonary function testing every three months, encompassing spirometry (\ac{FEV1}, \ac{FVC}) and \ac{DLCO}, should be integrated as a standard element of astronaut health surveillance during missions, facilitating the early identification of subclinical pulmonary fibrosis prior to the onset of irreversible functional impairment.

\subsection{Nickel Carcinogenic Risk: Speciation, Thresholds, and Clinical Management}

Nickel presents a qualitatively distinct risk profile from Cr and Co: it does not exceed the non-carcinogenic HI threshold (Ni \ac{HI}raw~=~0.04) but does exceed the acceptable cancer risk threshold, \ac{ILCR}~=~$9.2 \times 10^{-6}$ versus the $10^{-6}$ threshold, in the unfiltered scenario, with P(\ac{ILCR}~$>$~$10^{-6}$) $\approx$~100\% across Monte Carlo simulations under modelled exposure assumptions. This dichotomy arises from the structural difference between the non-carcinogenic and carcinogenic risk frameworks: the Reference Dose for Ni (0.02~mg~kg$^{-1}$~day$^{-1}$) is relatively permissive, while the Cancer Slope Factor of 0.91 reflects the \ac{IARC} Group 1 carcinogenicity classification of insoluble nickel compounds, particularly nickel subsulfide (Ni$_3$S$_2$) and nickel oxide (NiO), usually associated with nasal, sinus, and lung cancers \citep{linna2004, goodman2011}. In the martian basaltic context, Ni is expected to occur primarily as FeNi metal, Ni-bearing olivine, and Ni-containing sulfide phases, which are insoluble and chemically resistant, supporting the application of the insoluble Ni \ac{CSF} as a conservative but geochemically appropriate assumption \citep{yen2006, linna2004}.

At 95\% filtration, the Ni \ac{ILCR} decreases to $4.6 \times 10^{-7}$, below the $10^{-6}$ threshold, so establishing 95\% efficiency as the minimal engineering criteria for the tolerance of Ni cancer risk. In contrast to Cr, the carcinogenic risk associated with nickel can be entirely mitigated with a 95\% \ac{HEPA} filtering system, eliminating the need for further chemical treatments.

There is currently no medical intervention with proven effectiveness against nickel carcinogenicity. Diethyldithiocarbamate and several chelating substances have shown efficacy in mitigating acute nickel toxicity in animal models; however, they are not sanctioned for preventative application in humans \citep{bradberry1999, sprutt1978, nielsen1994}. The clinical management of nickel exposure primarily depends on engineering controls to reduce exposure, ensuring and sustaining a filtration efficiency of at least 95\%. This is complemented by pre-mission patch testing to identify nickel-sensitized individuals at increased risk of dermatological and immunological reactions, along with urinary nickel monitoring as a biomarker for systemic absorption during the mission.

\subsection{Sub-Threshold Elements Warranting Medical Surveillance: Iron, and Manganese}

Although Fe (\ac{HI}~=~0.43), Mn (\ac{HI}~=~0.05), and Al (\ac{HI}~=~0.04) do not surpass the non-carcinogenic \ac{HI} threshold of 1.0 in the unfiltered scenario, but their proximity for concern levels, particularly Fe, and the possibility of scenario extension necessitate systematic secondary monitoring. The absolute \ac{ADD} spectrum for Mars is dominated by iron (\ac{ADD}~=~$3.0 \times 10^{-1}$ mg~kg$^{-1}$~day$^{-1}$) by a significant margin, which is a direct result of the 14.1 weight\% Fe abundance in martian basaltic regolith. Fe \ac{HI} would approach or exceed 1.0 in scenarios of elevated exposure, such as an ingestion rate of 200~mg~day$^{-1}$ during \ac{EVA} suit integrity loss events or a multi-year conjunction-class mission that extends exposure duration. Hepatic siderosis, oxidative tissue injury, and cardiomyopathy are the results of chronic iron overload resulting from gastrointestinal Fe absorption \citep{chua1996, kremastinos2011}. At sustained elevations in body iron stores, haemochromatosis-like pathology is observed \citep{brissot2018}.

Manganese poses significant neurological risks. Chronic inhalation and ingestion of elevated manganese (Mn) is linked to manganism, a progressive extrapyramidal syndrome that clinically resembles idiopathic Parkinson's disease, due to preferential accumulation in the basal ganglia and dopaminergic neuronal damage caused by oxidative stress and mitochondrial dysfunction \citep{lucchini2023, peres2016}. Although the hazard index (\ac{HI}) for Mn is small for the specified 18-month mission parameters. For extended missions, especially in permanent settlement scenarios, Mn neurotoxicity must be explicitly modeled with mission-duration-specific hazard index estimates. It is advisable to conduct a pre-mission neurological baseline examination and frequent neurocognitive monitoring during all crewed missions on the martian surface.

\subsection{Limitations and Sources of Uncertainty}

Numerous significant limitations constrain the present findings and require attention in future research:

This study employs  the bulk silicate Mars model of \citet{taylor2013} as a proxy for the composition of surface regolith. The surface of Mars is recognized for its divergence from bulk composition due to aqueous weathering (enrichment of S and Cl in soils), aeolian sorting of mineral phases, and impact gardening. \ac{MSL} \ac{APXS} measurements at Gale Crater \citep{berger2020} indicate generally constant major element chemistry, although exhibit considerable local heterogeneity in trace elements; thus, the bulk composition serves as a globally-averaged estimate rather than a site-specific measurement.

Chromium speciation: Utilizing Cr(VI) toxicity metrics exemplifies a worst-case situation. Should future missions measure the Cr(III)/Cr(VI) ratio at the surface, the non-cancer hazard index and cancer incremental lifetime cancer risk could be significantly adjusted downward. This is arguably one of the important geochemical measurements for assessing human health risks on Mars.

Exclusion of perchlorate: Perchlorates are not modeled as the \ac{EPA} \ac{IRIS} does not offer a soil \ac{ADD}/\ac{RfD} framework for perchlorate in the format utilized for metals. This indicates a considerable disparity; persons subjected to perchlorate at doses of 0.5-1 weight\% would markedly above the threshold for thyroid dysfunction.

Synergistic toxicology: The \ac{HI} framework is entirely additive across different exposure pathways. Established synergies in metal toxicity, such as the co-exposure of Cr(VI) and Ni increasing lung cancer risk, or the interactions between concurrent heavy metal exposure and ionizing radiation, are not accounted for. These interactions would consistently elevate the actual health burden beyond the stated \ac{HI} levels.

\section{Recommendations}

\subsection{Engineering Recommendations}
Implement staged \ac{HEPA}/\ac{ULPA} ($\geq$99.97\% efficiency for $\geq$0.3~$\mu$m particles) combined with activated-carbon filtration in all pressurised habitats, airlocks, and \ac{EVA} suit life support systems. Design dedicated robotic or automated airlock cleaning and suit-doffing systems to minimize regolith tracking into habitats. Each \ac{EVA} cycle should include full quantitative decontamination. Establish a Mars Surface Dust Monitoring System to measure real-time inhalable Cr, Co, and Ni particle concentrations in habitat air, with automated alarm systems triggering enhanced filtration at elevated concentrations. Commission in situ Cr speciation measurements as a priority science objective for pre-crewed robotic missions, using portable X-ray absorption spectroscopy instrumentation or Raman spectroscopy.

\subsection{Medical and Operational Recommendations}

Implement pre-mission, in-mission, and post-mission biological monitoring for Cr and Co exposure, including urinary Cr(VI) and serum Co as biomarkers of internal dose. Consider prophylactic Vitamin C supplementation (500–1,000 mg/day) as a Cr(VI) reducing agent \citep{hadjiliadis2012}, with careful monitoring for kidney stone risk in the microgravity environment. Establish a martian regolith \acf{OEL} equivalent to the \ac{NIOSH} \ac{REL} for Cr(VI) of 0.2~$\mu$g/m$^3$, with interim \ac{OSHA} \ac{PEL} of 5~$\mu$g/m$^3$ as the absolute maximum. Include pulmonary function testing (such as \ac{FEV1}, \ac{FVC}, \ac{DLCO}) in the standard astronaut health surveillance protocol at 3-month intervals during and after Mars surface missions.

\subsection{Research Recommendations}

Prioritise toxicological studies of Mars regolith simulants, such as \ac{JSC}'s Mars-1A and \ac{MMS}, using in vitro pulmonary cell models to characterize combined metal + silica + perchlorate + \ac{UV}-activated \ac{ROS} toxicity. Develop a Mars-specific health risk framework integrating regolith metal toxicity, radiation dose (\ac{GCR} + \ac{SPE}), microgravity immunosuppression, and psychosocial stress as interactive, synergistic risk factors. Commission high-resolution \ac{GRS} reanalysis of Mars orbital data targeting Cr, Co, and Ni abundance maps to characterize regional variability and identify potential hotspot areas for habitat siting avoidance.

\section{Conclusions}

This study provides a comprehensive, multi-element, multi-pathway quantitative health risk assessment of martian regolith exposure for a 70~kg adult astronaut on an 18-month surface mission, as well as long-term carcinogenic risk, across four filtration scenarios. The principal geochemical health risk in martian regolith is chromium (Cr). Chromium (Cr) represents both a regulatory-threshold non-carcinogenic hazard and an increased carcinogenic risk, as evidenced by a Hazard Index (\ac{HI}) of 11.90 (raw) and an \acf{ILCR} of $5.8 \times 10^{-5}$, as well as a probability of P(\ac{HI}~$>$~1) at 100\% across 10,000 Monte Carlo simulations. A minimum filtering efficiency of 91.6\% is required to achieve \ac{HI}~$<$~1; a decrease in \ac{ILCR} below $10^{-6}$ requires an efficiency of 99\%. The martian mantle's near-chondritic, undifferentiated chromium abundance, in conjunction with the extremely oxidizing surface conditions on Mars, is the primary factor that promotes Cr(VI) speciation. 

Cobalt (Co) is the second most significant non-carcinogenic risk (\ac{HI}~=~5.08; P(\ac{HI}~$>$~1)~=~100\%). This is the result of a combination of the lowest oral \ac{RfD} and moderate geochemical enrichment relative to \ac{UCC}. This discovery, which is not present in the prior Mars dust health literature \citep{wang2025}, establishes Co as a priority in astronaut safety monitoring techniques, alongside Cr. In order to reduce Co \ac{HI} below the regulatory limit, a minimum filtering rate of 74.3\% is required. 

Nickel (Ni) exceeds the permissible cancer risk threshold (\ac{ILCR}~=~$9.2 \times 10^{-6} > 10^{-6}$) in the unfiltered scenario, with a probability of P(\ac{ILCR}~$> 10^{-6}$) of approximately 100\%, as indicated by Monte Carlo analysis. The Ni \ac{ILCR} decreases to $4.6 \times 10^{-7}$ at 95\% filtering, which is below the permissible level. Nickel's health risk profile is exclusively carcinogenic in the modeled scenarios, as it does not exceed the non-cancer \ac{HI} threshold (\ac{HI}~=~0.04). 

In the unfiltered scenario, all other elements have a \ac{HI} of less than 1 and an \ac{ILCR} of less than $10^{-6}$. The modelled exposure parameters do not present regulatory-level health risks for Be, As and Cd on Mars, in contrast to the evaluation by \citet{wang2025}. Our result is due to the fact that the elemental quantities on Mars are substantially lower than those in the \acf{UCC} of Earth. 

To concurrently satisfy the non-cancer hazard index (\ac{HI}) of less than 1 for chromium (Cr) and cobalt (Co) and the \ac{ILCR} of less than $10^{-6}$ for nickel (Ni), it is recommended that habitat and \ac{EVA} suits maintain a minimum filtration level of 95\% efficiency. This criterion can be technically satisfied with \ac{HEPA} systems; however, additional surface disinfection is required to reduce contamination through the primary ingestion channel. Despite 95\% filtration, a residual chromium carcinogenic risk (\ac{ILCR}~=~$2.9 \times 10^{-6}$) persists, necessitating medicinal interventions, such as Vitamin C ingestion as a Cr (VI) lowering agent, as a secondary treatment. 

In conclusion, this study demonstrates that the geochemical health hazard profile of martian regolith is qualitatively distinct from that of terrestrial analogues. This distinction is attributed to the primitive, undifferentiated nature of the martian crust, rather than anthropogenic contamination. The effective management of Cr and Co non-carcinogenic hazards, as well as Cr and Ni carcinogenic risks, can be achieved through a combined engineering and medical approach. This approach includes the implementation of \ac{HEPA}/\ac{ULPA} air filtration at $\geq$95\% efficiency, systematic airlock decontamination protocols, biological monitoring of urinary Cr(VI) and serum Co, and Vitamin C supplementation. The implementation of these measures in the design of the surface habitat on Mars should be regarded as a mandatory engineering requirement, rather than a precautionary option.

\section*{Acknowledgements}
\begin{acknowledgments}
We thank the reviewers for their constructive comments on this paper. Author MA was funded by the University of Idaho.
\end{acknowledgments}

\appendix
\section{Acronyms}
\begin{singlespace}
\begin{acronym}
    \acro{ABSd}{Dermal Absorption fraction}
    \acro{ADD}{Average Daily Dose}
    \acro{AF}{soil-to-skin Adherence Factor}
    \acro{APXS}{Alpha Particle X-ray Spectrometer}
    \acro{ATnc}{Average Time for non-cancer risk}
    \acro{ATc}{Time for lifetime cancer risk}
    \acro{ATSDR}{Agency for Toxic Substances and Disease Registry of the \ac{CDC}}
    \acro{CDC}{Centers for Disease control of the \ac{US}}
    \acro{CSF}{Cancer Slope Factor}
    \acro{CV}{Coefficients of Variation }
    \acro{ED}{Exposure Duration}
    \acro{EnFac}{Enrichment Factor}
    \acro{ExFreq}{Exposure Frequency}
    \acro{EPA}{U.S.\ Environmental Protection Agency }
    \acro{ERF}{Ecological Risk Factor}
    \acro{EVA}{Extravehicular Activity}
    \acro{FEV1}{Forced Expiratory Volume in 1~second}
    \acro{FVC}{Forced Vital Capacity}
    \acro{DLCO}{Diffusing Capacity of the Lungs for Carbon Monoxide}
    \acro{GCR}{Galactic Cosmic Ray}
    \acro{GRS}{Gamma-Ray Spectrometer}
    \acro{HEPA}{High-Efficiency Particulate Air}
    \acro{HI}{Hazard Index}
    \acro{HQ}{Hazard Quotient}
    \acro{HHRA}{Human Health Risk Assessment}
    \acro{HMLD}{Hard Metal Lung Disease}
    \acro{IARC}{International Agency for Research on Cancer}
    \acro{ILCR}{Incremental Lifetime Cancer Risk}
    \acro{IR}{Inhalation Rate}
    \acro{IUR}{Inhalation Unit Risk}
    \acro{IRIS}{Integrated Risk Information System}
    \acro{JSC}{Johnson Space Center}
    \acro{LILE}{Large-Ion Lithophile Elements}
    \acro{MDF}{Mars Dust Enhancement Factor}
    \acro{MMS}{Mojave Mars Simulant}
    \acro{MSL}{Mars Science Laboratory}
    \acro{NASA}{National Aeronautics and Space Administration}
    \acro{NIOSH}{\ac{US} National Institute for Occupational Safety and Health}
    \acro{OEL}{Occupational Exposure Limit}
    \acro{OSHA}{Occupational Safety and Health Administration}
    \acro{PEL}{Permissible Exposure Limit}
    \acro{RAD}{Radiation Assessment Detector}
    \acro{RAGS}{Risk Assessment Guidance for Superfund}
    \acro{REE}{Rare Earth Elements}
    \acro{REL}{Recommended Exposure Limit}
    \acro{RfD}{Chronic Oral Reference Dose}
    \acro{RIng}{Oral INGestion Rate}
    \acro{ROS}{Reactive Oxygen Species}
    \acro{SA}{Exposed Skin Surface Area}
    \acro{SPE}{Solar Particle Event}
    \acro{TRF}{Toxic Response Factor}
    \acro{ULPA}{Ultra-Low Penetration Air}
    \acro{UCC}{Upper Continental Crust}
    \acro{US}{United States of America}
    \acro{UV}{Ultra-Violet}
    \acro{XANES}{X-ray Absorption Near-Edge Structure}
\end{acronym}
\end{singlespace}

\newpage
\bibliography{healthHazards}{}
\bibliographystyle{psj}

\end{document}